\documentclass[nofootinbib,showpacs,preprintnumbers,amsmath,amssymb,floatfix]
{revtex4-2}
\usepackage{graphicx}

\begin{document}


\title{Parton helicities at arbitrary $x$ and $Q^2$ in double-logarithmic approximation}


\author{B.I.~Ermolaev}
\affiliation{Ioffe Institute, 194021
 St.Petersburg, Russia}

\begin{abstract}
Description of spin-dependent hadronic processes at high energies in terms of parton helicities is a both 
effective and technically convenient means. 
In the present paper, we obtain explicit expressions for the parton helicities 
when either Collinear or $KT$ forms of QCD Factorization are used.  
Starting our studies with calculation of the helicities   
in the double-logarithmic approximation (DLA) in the region of small $x$ and large $Q^2$, 
we generalize the results in order to obtain formulae  valid at 
arbitrary $x$ and $Q^2$. We argue against using Collinear Factorization, when the parton  orbital angular momenta 
are accounted for, and prove that $KT$ Factorization should be used instead. 
We also consider in detail the small-$x$ asymptotics of the parton helicities, compare them with the DGLAP-asymptotics 
in LO,NLO, etc and prove that the DGLAP asymptotics are less singular at small $x$ than the 
Regge asymptotics
\end{abstract}


\maketitle

\section{Introduction}

Parton helicities correspond to longitudinal components of the parton spins 
when hadron masses are neglected and by this reason the helicities 
have often been used for description of spin-dependent hadronic reactions  at high energies. A new  wave of interest to 
parton helicities was initiated recently by Yu.V.~Kovchegov and his colleagues who investigated  the proton spin at high energies. 
They suggested in Refs.~\cite{kovchfirst}-\cite{smalldis} 
 a new method, namely KPSCTT evolution equations, to calculate the quark and gluon helicities. Their calculations were done mostly with double-logarithmic (DL) 
 accuracy though some of the papers included 
 single-logarithmic contributions as well. The obtained results were represented in the form of the small-$x$ asymptotics.  
 To make sure that the KPSCTT was correct, the small-$x$ asymptotics of the parton helicities were compared with the asymptotics of the spin-dependent structure 
 function $g_1$ calculated in Ref.~\cite{bers} by other means. The comparison led to the full agreement between those asymptotics. 
After that, the KPSCTT method was applied  to calculations of the proton spin at high energies in Refs.~\cite{agr}-\cite{global} 
and to  scattering of polarized protons in Ref.~\cite{kovscatt}. \\
 
 Recognizing that description of the nucleon spin at high energies in terms of parton helicities was the optimal means for 
 studying the proton spin problem, we 
investigated this problem  in Ref.~\cite{espin} in terms of the parton helicities with DL accuracy. Both the quark and gluon 
helicities (which we denote $h_{q,g}$ throughout the present paper)  
depend on $x = Q^2/(2pq)$ and $Q^2$ (with $q$ and $p$ being momenta of the parton and the proton respectively). 
 We skipped tracing the $Q^2$-dependence in Ref.~\cite{espin}
 as we followed the pattern of the RHIC experiments\cite{rhic1,rhic2} where $Q^2$ was fixed at $10$~GeV$^2$. 
 This approximation allowed us to apply the results on $h_{q,g}$ 
obtained in Ref.~\cite{egtg1s} ( see also the overview Ref.~\cite{egtg1sum}). 

However, accounting for the $Q^2$-dependence is mandatory in order to get a complete of the parton helicities. Moreover, 
the formulae for $h_{q,g}$ of Refs.~\cite{egtg1s,egtg1sum}  can be used in combination with 
Collinear Factorization only, they are absolutely incompatible with $KT$ Factorization. 
Nevertheless,
there are situations where the use of $KT$ Factorization can be beneficiary and even unavoidable. 
It makes necessary deriving expressions for the parton helicities in the form compatible with $KT$ Factorization. 
Obtaining  such  
expressions for the parton helicities that account for both $x$- and $Q^2$-dependence and also are  
compatible with both Collinear and KT Factorizations 
is the main subject of the present paper. First we calculate them
in the DL approximation (DLA) at small $x$ and large $Q^2$, and then, engaging  the DGLAP 
formulae, construct interpolation 
formulae valid at arbitrary $x$ and $Q^2$. 

Our calculations 
are done with composing and solving appropriate Infra-Red Evolution 
Equations (IREE)\footnote{Detailed explanations on constructing and solving IREEs in the DIS context can be found in the overview\cite{egtg1sum} whilst a brief description 
of this method is given in Sect.~II.}. 
The key point here is that DL contributions to amplitudes (i.e. scattering amplitudes, structure functions, form factors, etc.) of QCD processes arrive from the integration regions where some of the  
virtual partons are almost on-shell.  In this case, 
 the amplitudes become IR-singular and they need to be regulated by IR cut-offs $\mu$ which are primarily free parameters and specified a posteriori. Studying evolution of the amplitudes 
with respect to $\mu$ is the essence of the IREE approach.  
Constructing IREEs exploits factorization of DL contributions of virtual particles with minimal transverse momentum $k_{\perp}$ 
even if their energies are not small, which was first noticed in the QED context by V.N.~Gribov in Ref.~\cite{g} and then extended to non-Abelian 
theories by L.N.~Lipatov\cite{l1,l2,kl1,kl2}. IREEs in the QCD context for $2 \to 2$ processes were suggested in Ref.~\cite{kl1,kl2} 
whilst IREEs for inelastic $2 \to 2 + n$ processes were  suggested in Refs.~\cite{efl,el}. The IREE approach proved to be much simpler  
 than preceding tools even in the relatively simple case of elastic processes (cf. for example Refs.~\cite{kl1,kl2} and \cite{ggfl}), 
 not to mention inelastic reactions, see Refs.~\cite{efl,el}. 
 
In contrast to Ref.~\cite{espin}, we consider in more detail technique of constructing IREEs for the helicities and dwell upon 
the role of the 
$t$-channel color octet amplitudes. As a result, we obtain the expressions for $h_{q,g}(x,Q^2)$ which can be used in the framework of 
Collinear Factorization and also more complex expressions compatible with KT factorization. We argue that it is 
KT Factorization that is supposed to be used when parton orbital angular momenta (OAM) are accounted for.   

After obtaining explicit expressions for $h_{q,g}(x,Q^2)$,  we consider in detail calculating their small-$x$ asymptotics, 
applying the Saddle-Point method to their parent expressions. 
By doing so, we demonstrate  that despite $h_q(x,Q^2)$, $h_g(x,Q^2)$ and 
$g_1(x,Q^2)$ in DLA are represented by different formulae, 
their asymptotics are identical.  
Comparing these asymptotics with the ones obtained in the framework of N$^n$LO DGLAP, we prove that the 
DGLAP\cite{dglap} asymptotics in LO, NLO, NNLO, etc do not acquire the Regge form. 
Then we examine the impact 
of the involved fits on the small-$x$ behaviour of the helicities and discuss a problem of "false intercept" i.e. of the intercept supposedly dependent on 
$Q^2$.   \\
 
Our paper is organized as follows: in Sect.~II we briefly remind basic features of the IREE method and introduce our basic definitions.  
We present there a brief prescription for constructing IREEs in general. 
In order to account for both 
 perturbative and non-perturbative contributions to the parton helicities, we introduce in Sect.~III their representation 
 throughout QCD Factorization and discuss involved mass scales. The kinematic regions where formulae for $h_{q,g}$ are 
 different are introduced in Sect.~IV. IREEs for perturbative contributions to the perturbative components of 
 $h_{q,g}$, where all external partons are off-shell,  are derived in Sec.~V. A simpler case involving 
 both off-shell and on-shell external partons is considered in Sect.~VI 
 whereas the case of only on-shell partons is for Sect.~VII.  In contrast to the IREEs for the DIS structure function $g_1$, 
 IREEs for the parton helicities involve the $t$-channel color octet contributions from the starting point. 
 We discuss the situation with the color octets 
 in Sect.~VIII.  Explicit DL expressions for  $h_{q,g}$ in the context of both Collinear and KT Factorization    
 are presented in Sect.~IX. Besides, we discuss there using Orbital Angular Momenta for 
   description of nucleon spins. Sect.~X is for extending these expressions to arbitrary $x$ and $Q^2$ so as to 
   construct interpolation formulae for $h_{q,g}(x,Q^2)$ valid at arbitrary $x$ and $Q^2$.  
 Sect.~XI contains detailed consideration of small-$x$ asymptotics 
 of $h_{q,g}$. Finally, Sect.~XII is for our concluding remarks.     

\section{Preliminary notes}

\subsection{List of our basic notations}

To make reading easier, we list below basic notations that we use throughout the paper:\\

$h_{q,g}\left(x, Q^2\right)$  denote the parton helicities, they are defined in Eqs.~(\ref{factcol},\ref{factkt}). \\

$\Phi_{q,g}$ are initial parton distributions in Collinear Factorization, introduced in Eq.~(\ref{factcol}).\\

$\varphi_{q,g}$ are  related to $\Phi_{q,g}$ by Eq.~(\ref{mellin}).\\

$\Phi^{KT}_{q,g}$ are initial parton distributions in KT Factorization, introduced in Eq.~\ref{factkt}).\\

$\varphi^{KT}_{q,g}$ are  related to $\Phi^{KT}_{q,g}$ by Eq.~(\ref{mellin}).\\

$A_{ij}\left(w, Q^2, k^2_{\perp}\right)$ denotes totally off-shell parton-parton amplitudes, introduced in Eq.~(\ref{mellin}). \\

$M_{ij}\left(w, Q^2, k^2_{\perp}\right)$ denotes imaginary parts of amplitudes $A_{ij}\left(w, Q^2, k^2_{\perp}\right)$, 
introduced in Eq.~(\ref{ima}).\\

$F_{ij}\left(\omega, Q^2,k^2_{\perp}\right) $ are related to amplitudes $A_{ij}\left(w, Q^2, k^2_{\perp}\right)$ 
by Eq.~(\ref{mellin}). \\

$A^{\prime}_{ij}\left(w, Q^2\right)$ denotes partly off-shell parton-parton amplitudes, introduced in 
Eq.~(\ref{mmprime}).\\

$M^{\prime}_{ij}\left(w, Q^2\right)$ denotes imaginary parts of amplitudes $A_{ij}\left(w, Q^2, k^2_{\perp}\right)$, 
introduced in 
Eq.~(\ref{mmprime}).\\

$F^{\prime}_{ij}\left(\omega, Q^2\right) $ are related to amplitudes $A_{ij}\left(w, Q^2\right)$ 
by Eq.~(\ref{mellin}).\\

$A^{\prime \prime}_{ij}\left(w\right)$ denotes on-shell parton-parton amplitudes, introduced in Eq.~(\ref{am2prime}).\\

$M^{\prime \prime}_{ij}\left(w\right)$ denotes imaginary parts of amplitudes $A^{\prime \prime}_{ij}\left(w\right)$, 
introduced in Eq.~(\ref{am2prime}).\\

$f_{ij}\left(\omega\right) $ are related to amplitudes $A^{\prime \prime}_{ij}\left(w\right)$ 
by Eq.~(\ref{mellin}).\\

\subsection{Mellin transform}

Throughout the paper we will use the Mellin transform which 
is defined as follows:

 \begin{equation}\label{mellin}
 A_{ij}\left(w, Q^2,k^2_{\perp}\right) = \int_{-\imath \infty}^{\imath \infty} \frac{d \omega}{2 \pi \imath} 
 \left(w/\mu^2\right)^{ \omega}
  F_{ij}\left(\omega, Q^2,k^2_{\perp}\right),
 \end{equation}
 where $w = 2pq$ and $\mu$ is a mass scale. 
 It is often associated with  the factorization scale.  We consider its value in the next Sect.  
 The integration line in Eq.~(\ref{mellin}) runs to the right of the
 rightmost singularity of $F_{ij}\left(\omega, y\right)$. 
  It is convenient to consider 
 $A_{ij}\left(x, Q^2,k^2_{\perp}\right)$,
  using the logarithmic notations:

 \begin{equation}\label{mellinlog}
 A_{ij}\left(w, Q^2,k^2_{\perp}\right) 
  = \int_{-\imath \infty}^{\imath \infty} \frac{d \omega}{2 \pi \imath} e^{\omega \rho}
F_{ij}\left(\omega, y_1,y_2\right),
 \end{equation}
 where the logarithmic variables $\rho,\xi, y$ defined as follows:

 \begin{equation}\label{rhoy12}
 \rho = \ln \left(w/\mu^2\right), ~~y_1 = \ln \left(Q^2/\mu^2\right),y_2 = \ln \left(k_{\perp}^2/\mu^2\right)~~\xi = \rho - y_1 = \ln \left(1/x\right).
 \end{equation}

Definition of $\rho$ in Eq.~(\ref{rhoy12}) does not respect analytical properties of $w$-dependence of an  
amplitude $A_{ij} (w)$.
Indeed, $A_{ij} (w)$ should have a cut at positive $s = (p+q)^2 \approx w$. It means that $\ln w$ should be replaced by 
$\ln \left(s\;e^{- \imath \pi}\right) = \ln w - \imath \pi$ and therefore 

\begin{equation}\label{aser}
A_{ij} \left(\ln w - \imath \pi\right)  \approx A_{ij} (\rho) - \imath \pi \frac{d A_{ij}}{ d \rho},
\end{equation}
i.e. 

\begin{equation}\label{ima}
M_{ij}   \approx - \pi \frac{d A_{ij}}{ d \rho}.
\end{equation}

Combining Eqs.~(\ref{ima}) and (\ref{mellinlog}), obtain 

 \begin{equation}\label{mellinm}
 M_{ij}\left(w,y_1,y_2\right) 
  = - \pi \frac{d}{d \rho}\int_{-\imath \infty}^{\imath \infty} \frac{d \omega}{2 \pi \imath} e^{\omega \rho}
F_{ij}\left(\omega\right) = - \pi\;
\int_{-\imath \infty}^{\imath \infty} \frac{d \omega}{2 \pi \imath} e^{\omega \rho}\;
\omega \; F_{ij}\left(\omega,y_1,y_2\right).  
\end{equation}
In what follows we will drop the factor $\pi$, assuming that it can be attributed to the parton 
distributions $\Phi_{q,g}$.  We will use the same transform to relate $M^{\prime}_{ij}$ to $F^{\prime}_{ij}$ 
and $M^{\prime \prime}_{ij}$ to $f_{ij}(\omega)$. 

\subsection{General remark on the IREE technology}

Technology of constructing IREEs is universal for all QCD reactions. We briefly remind it below, using as an example 
a relatively simple case of $2 \to 2$ scattering of partons. We denote $A$ the scattering amplitude of this process. 
 IREE for amplitude $A$ is depicted in the diagram form in Fig.~1. 
 DL contributions to $A$ arrive from the kinematics where transverse momenta $k_{j \perp}$, $j = 1,..$ 
of virtual partons are strongly ordered, so 
there always can be found a virtual parton with minimal $k_{\perp}$ (softest parton) although its energy can be large.  
DL contribution of such parton can be factorized. 
\begin{figure}\label{helQfig1}
\includegraphics[width=.6\textwidth]{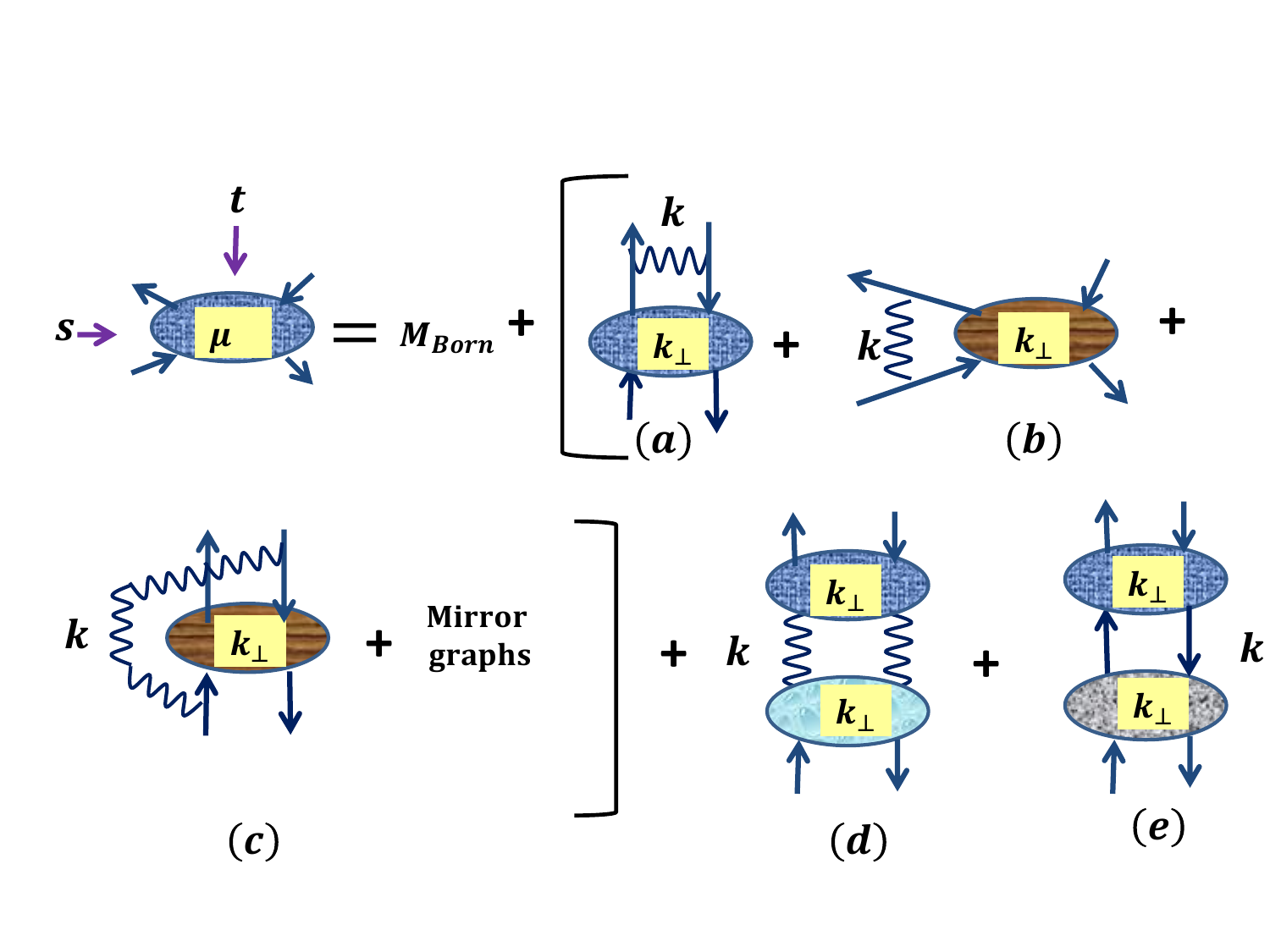}
\caption{\label{helQfig1}
Constructing IREE for amplitude $A$ of $2 \to 2$- parton scattering, using  factorization of the softest partons with momentum $k$. 
Letters $s,t$ in the l.h.s. denote the standard Mandelstam variables. The first term in the r.h.s. stands for the Born contribution. 
The blobs denote that radiative corrections in DLA 
are accounted for. The blobs on graphs (b,c) (marked with horizontal strips) are $t$-channel color octets while the other blobs are color singlets.  The letters on the blobs stand for IR 
cut-offs. Graphs (a,b,c) correspond to factorization of non-ladder partons (Option (i)) whilst factorization of ladder ones (Option (ii)) is shown on graphs (d) and (e). }
\end{figure}
There are two options to factorize the softest parton:\\ 
\textbf{Option (i)} The softest parton is a non-ladder gluon\footnote{We imply the Feynman gauge.}. Its factorizing means that its propagator is attached to pairs of 
the external 
lines in every possible way as shown on graphs (a,b,c)in Fig.~1 (mirror graphs  are implied)  
whereas $k_{\perp}$ acts as a new IR cut-offs for  
momenta of all other 
virtual partons in the blobs.  Note that graph (a) does not yield DL contribution at $t = 0$, so we will drop it in what follows. 
The blobs on graphs (b,c) correspond to the $t$-channel colour octets. Indeed, the blob in the l.h.s. is the color singlet but after 
emission of the softest gluon, the blob corresponds to the sum of non-singlet $SU(3)$-representations 
in the $t$-channel. The simplest non-singlet representation is the octet. It can be described in terms of gluon 
wave functions while the higher representations require introducing other, exotic fields. By this reason we consider the 
octet contributions only. Introducing the octet and singlet projection operators and applying them in turn both to graphs (a,b,c) in 
Fig.~1 and to the blobs on these graphs 
demonstrates that the contributions of the singlet blobs are zeros which leaves us with the octet blobs only, see details e.g. in 
Ref.~\cite{kl}.  \\
\textbf{Option (ii)} The softest parton ( quark or gluon) is a ladder parton. Result of its factorizing is that the primary amplitude is represented by convolutions (d) and (e) 
 of two $2 \to 2$-amplitudes whereas $k_{\perp}$ acts again as a new IR cut-off  for all other virtual partons. 
 All blobs in these convolutions are color singlets. Note that 
the intermediate states, with the number of $t$-channel partons greater than 2, do not bring DL contributions.  
Integration over $k_{\perp}$ involves $\mu$ as a lowest limit. Value of $\mu$ is primarily arbitrary save the obvious requirement 
$\mu > \Lambda_{QCD}$. Specifying $\mu$ is discussed in the next Sect.

Applying the standard Feynman rules to the graphs in Fig.~1 leads to converting this IREE to analytic form. The most important difference 
between IREEs and other equations of  the Bethe-Salpeter type is that the blobs/kernels in Fig.~1 do not depend on longitudinal components 
of $k$ though they depend on $k_{\perp}$ in the sense that $k_{\perp}$ acts a new IR cut-off for momenta of other virtual partons.  Finally, notice that the convolutions (d,e) in Fig.~1 
look much simpler in the $\omega$-space after the Mellin transform has been used. 
  This technology was used to describe  many various QED, QCD and Electroweak processes in DLA.   We apply 
 it below to calculate the parton helicities.

\section{Parton helicities within the framework of  QCD Factorization }

According to the QCD Factorization concept, the parton helicities $h_i \left(x, Q^2\right)$,   can be represented as
convolutions of the perturbative components $M_{ij}$ (i,j = q,g) and polarized
parton distributions $\Phi_j$. 
In the framework of Collinear Factorization  depend on 
two arguments: $M_{ij}= M_{ij}\left(x, Q^2\right)$ $M_{ij}$ while virtualities of the initial  
partons are $\approx \mu^2$, where $\mu$ is the factorization scale. It also plays the role of mass shell 
for the partons, so we will address  as partly off-shell amplitudes and denote 
$M^{\prime}_{ij}\left(x, Q^2\right)$.  
Therefore 

\begin{eqnarray}\label{factcol}
h_q \left(x, Q^2\right) &=& M^{\prime}_{qq}\left(x, Q^2\right)\otimes \Phi_q (x) +  
M^{\prime}_{qg}\left(x, Q^2\right)\otimes \Phi_g (x),
\\ \nonumber
h_g \left(x, Q^2\right) &=& M^{\prime}_{gq}\left(x, Q^2\right)\otimes \Phi_q (x) +  
M^{\prime}_{gg}\left(x, Q^2\right)\otimes \Phi_g (x).
\end{eqnarray}

  In contrast, $K_T$ Factorization involves dependence of $M_{ij}$ on one more argument: 
$M_{ij} = M_{ij}\left(x, Q^2,k^2_{\perp}\right)$, so the initial parton distributions $\Phi^{KT}_{q,g}$ 
depend on $x$ and $k^2_{\perp}$: 

\begin{eqnarray}\label{factkt}
h_q \left(x, Q^2\right) &=& M^{KT}_{qq}\left(x, Q^2, k^2_{\perp}\right)\otimes \Phi^{KT}_q (x,k^2_{\perp}) +  
M^{KT}_{qg}\left(x, Q^2,k^2_{\perp}\right)\otimes \Phi^{KT}_g (x,k^2_{\perp}),
\\ \nonumber
h_g \left(x, Q^2\right) &=& M^{KT}_{gq}\left(x, Q^2,k^2_{\perp}\right)\otimes \Phi^{KT}_q (x,k^2_{\perp}) +  
M^{KT}_{gg}\left(x, Q^2,k^2_{\perp}\right)\otimes \Phi^{KT}_g (x,k^2_{\perp}).
\end{eqnarray}

In what follows we will name amplitudes depending on three arguments, i.e. the ones like 
$A^{KT}_{ij}, M^{KT}_{ij}$, totally off-shell amplitudes. The partly off-shell amplitudes   
$A^{\prime}_{ij}$ and $M^{\prime}_{ij}$  are related to the totally off-shell 
amplitudes $A_{ij},M_{ij}$ as follows: 

\begin{equation}\label{mmprime}
 A_{ij}\left(x, Q^2,k^2_{\perp}\right)|_{k^2_{\perp} = \mu^2}  = A^{\prime}_{ij}\left(x, Q^2\right), ~~
 M_{ij}\left(x, Q^2,k^2_{\perp}\right)|_{k^2_{\perp} = \mu^2}  = M^{\prime}_{ij}\left(x, Q^2\right).
\end{equation}

On-shell parton-parton amplitudes $A^{\prime \prime}_{ij}(x)$ and their imaginary parts $M^{\prime \prime}_{ij}(x)$ 
are defined similarly: 

\begin{equation}\label{am2prime}
A^{\prime}_{ij}\left(x, Q^2\right)|_{Q^2 = \mu^2} = A^{\prime \prime}_{ij}(x), ~~
M^{\prime}_{ij}\left(x, Q^2\right)|_{Q^2 = \mu^2} = M^{\prime \prime}_{ij}(x).
\end{equation}

The convolution symbol $\otimes$ in Eqs.~(\ref{factcol},\ref{factkt}) implies integrations over momenta of intermediate $t$-channel partons. The number of 
involved  integrations depends on the specific form of Factorization.  
In Collinear Factorization $\otimes$ means the integrations over the longitudinal components of the intermediate parton momenta 
and $K_T$ Factorization involves in addition integration over $k_{\perp}$. 
Eqs.~(\ref{factcol},\ref{factkt}) are illustrated in Fig.~2, where the $s$-cuts  of each graph are implied.
\begin{figure}\label{helQfig3}
\includegraphics[width=.6\textwidth]{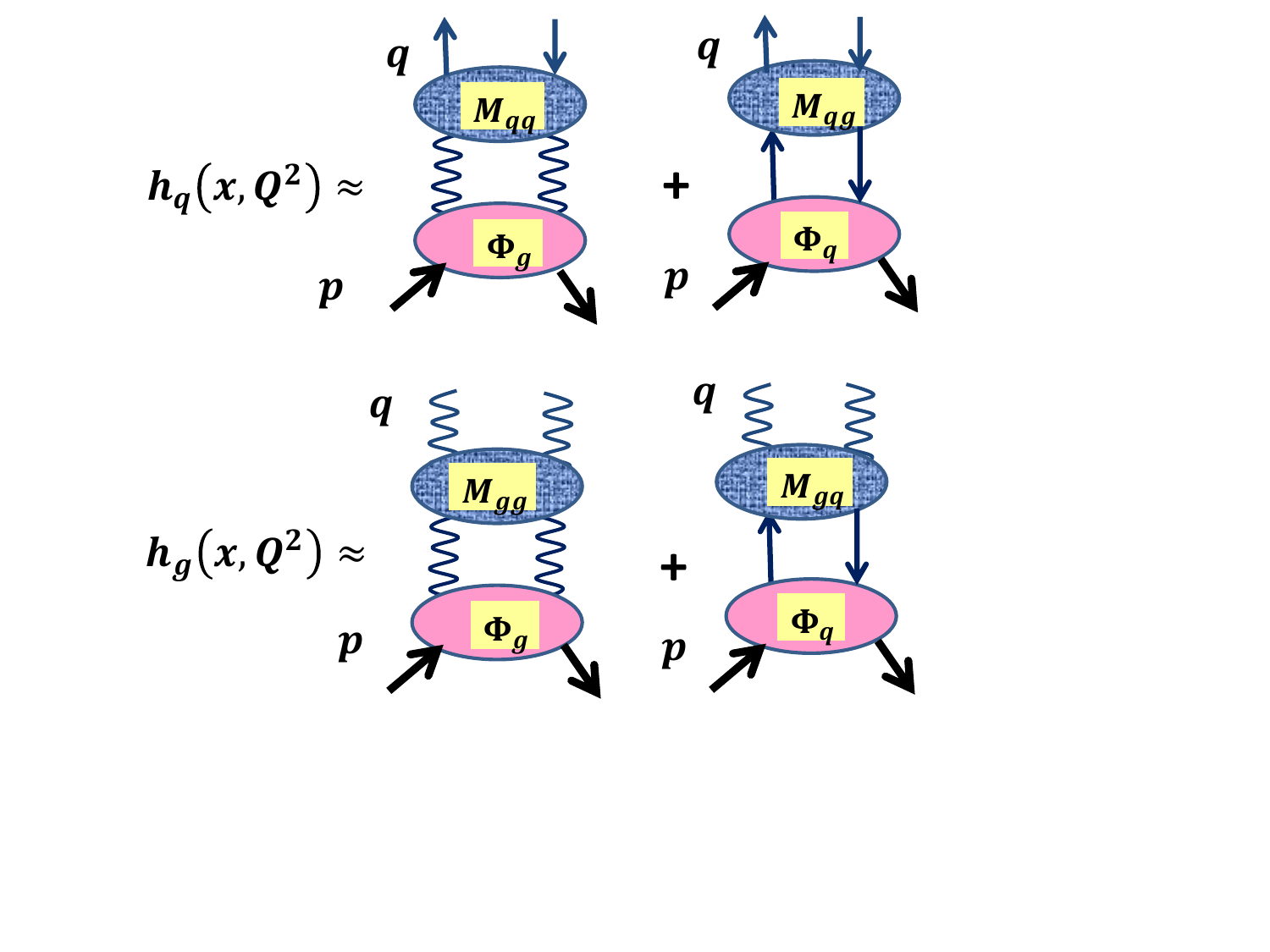}
\caption{\label{helQfig3}
Representation of the parton helicities through convolutions of perturbative (upper blobs) and non-perturbative 
(lower blobs) components in any of the forms of QCD Factorization. Waved lines denote gluons with momenta $q$, the thick straight lines 
attached to the lowest blobs correspond to the hadron with momentum $p$, the other 
straight lines attached to the upper blobs denote quarks with momenta $q$. The cuts with respect to $s = (p + q)^2$ on each graph are implied.}
\end{figure}
 In general,
the number of the intermediate $t$-channel partons in convolutions is not fixed, it depends on the type of
the parton-parton collisions under consideration. We account for the single-parton collisions,
which corresponds to two-parton  intermediate
states, as depicted in Fig.~2. Fortunately, the single-parton collision scenario is 
perfectly consistent with applying DLA. 

\subsection{Remark on mass scales}

Strictly speaking, theory of polarized DIS involves several mass scales which sometimes 
are used in the literature:\\
\textbf{(i)} Factorization scale $\mu_F$;\\
\textbf{(ii)} IR cut-off $\mu_{IR}$ for regulating IR singularities related to DL contributions to perturbative components of the parton 
helicities;\\
\textbf{(iii)} Starting point $Q^2_0$ of the $Q^2$-evolution; \\
\textbf{ (iv)} The mass scale $\mu$ in the Mellin transform (\ref{mellin}).\\
The IREE approach involves the IR cut-off in the transverse space, so $\mu_{IR}$ is the minimal transferred momentum 
and because of that it plays the role of the mass shell for the partons. Its value was estimated in Ref.~\cite{egtg1s,egtg1sum} on basis of Principle of Minimal Sensitivity\cite{pms}: $\mu_{IR} \approx 1$~GeV. Throughout the paper we will denote it $\mu$. 
For the sake of simplicity, we suggest that values of the factorization scale 
and  $Q_0$ are also $\approx \mu$ although this simplification is not principal for both our reasoning and our results.

\section{Kinematic regions for integration of the parton helicities}

Both $M_{ij}$ and the parent amplitudes $A_{ij}$ are given by different formulae,   
depending on $x$ and $Q^2$. Below we present these regions.  
The whole kinematic region $D_{tot}$ where the  variables $x$ and $Q^2$ run is 

\begin{equation}\label{regtot}
D_{tot}:  0 < x < 1, ~~ 0 \leq Q^2 < w.
\end{equation}

We divide $D_{tot}$ into four regions:

\begin{equation}\label{dabcd}
D_{tot} = D_A \oplus D_B  \oplus D_C  \oplus D_D. 
\end{equation}

First of all, there is the "hard" region $D_A$:

\begin{equation}\label{rega}
D_A :~~  x \sim 1, ~~Q^2 \gg \mu^2.
\end{equation}
As is well-known, DGLAP was especially constructed for operating in this region. 
In what follows, we will address $D_A$ as the DGLAP kinematic region. 
The next is region $D_B$, where DL contributions are leading:

\begin{equation}\label{regb}
D_B :~~  x \ll 1, ~~Q^2 \gg \mu^2.
\end{equation}
In what follows, we will address $D_B$ as the DL kinematic region.
Both $D_A$ and $D_B$ are the large-$Q^2$ regions where $Q^2 > \mu^2$. In contrast to them, there are also  
the small-$Q^2$ regions $D_C$ and $D_D$ in $D_{tot}$, 
where $Q^2 \leq \mu^2$:

\begin{equation}\label{regc}
D_C :~~  x \ll 1, ~~Q^2 \leq \mu^2,
\end{equation}

\begin{equation}\label{regd}
D_D :~~  x \sim 1, ~~Q^2 \leq \mu^2.
\end{equation}

We start with calculating the totally off-shell and partly off-shell amplitudes in the region $D_B$ and then move to considering the other regions. The technical means 
tat we will use in region $D_B$ is constructing and solving IREEs. However, there are two subregions of $D_B$, where IREEs for the totally off-shell amplitudes $A_{ij} (w, Q^2, k^2_{\perp})$ are different.  
%
These subregions are: \\

\textbf{(i)} Moderate-Virtual (MV) kinematics, where 

\begin{equation}\label{mvkin}
Q^2k^2_{\perp} < w \mu^2.
\end{equation}

Eq.~(\ref{mvkin}) can also be written as follows: 

\begin{equation}\label{mvkinx}
k^2_{\perp} < \frac{\mu^2}{x}. 
\end{equation}

In what follows we will denote $D^{MV}$ the subregion (\ref{mvkin}) of $D_B$ and keep the notations 
$A_{ij}(w,Q^2, k^2_{\perp})$ for the off-shell amplitudes in $D^{MV}$. \\

\textbf{(ii)} Deeply-Virtual (DV) kinematics, where 

\begin{equation}\label{dvkin}
Q^2k^2_{\perp} > w \mu^2, 
\end{equation}
i.e. 

\begin{equation}\label{dvkinx}
k^2_{\perp} > \frac{\mu^2}{x}. 
\end{equation}

Throughout the paper we will keep the notation $D^{DV}$ for this subregion and denote $A^{DV}_{ij}(w,Q^2, k^2_{\perp})$ the off-shell amplitudes in the region (\ref{dvkin}). 

\section{Totally off-shell amplitudes in region $D_B$ }

In the present Sect. we first 
demonstrate how to apply the general prescription for constructing IREEs presented in Sect.~IIC to constructing IREEs for amplitudes $A_{ij}$ 
and solve the obtained equations. Then, we move to the totally off-shell amplitudes in DV kinematics. \\

\subsection{Totally off-shell  amplitudes in region $D^{MV}$}

We demonstrate below how to apply the general prescription depicted in Fig.~1 to constructing IREEs for amplitudes $A_{ij}(w,Q^2, k^2_{\perp})$. 
 It is convenient to write IREEs in the 
differential form,
applying operator $-\mu^2 d/d \mu^2$ to all graphs in Fig.~1.
Also, it is convenient to write the IREEs in the Mellin representation because the convolutions corresponding to 
graphs (d,e) in Fig.~1 look much simple in the $\omega$-space. We denote $F_{ij} (\omega,y_1,y_2)$ the 
Mellin amplitudes corresponding to  $A_{ij}(w,Q^2, k^2_{\perp})$. As a result, we obtain the following IREE:


\begin{equation}\label{eqfmv}
\left[\frac{\partial }{\partial y_1} +  \frac{\partial}{\partial y_2} 
+ \omega \right] F_{ij} (\omega,y_1,y_2)   = 
\frac{1}{8 \pi^2}F^{\prime}_{in} (\omega, y_1) F^{\prime}_{nj} (\omega, y_2) + \frac{1}{8 \pi^2} \widetilde{T}_{ij} (\omega,y_1,y_2),
\end{equation}
where $i,j,n = q,g$ and summation over $n$ is implied. 
The l.h.s. of Eq.~(\ref{eqfmv}) corresponds to differentiation of l.h.s. of the IREE in Fig.~1. It contains two derivatives 
while the last term corresponds to differentiation 
of the factor $e^{\omega \rho}$ in Eq.~(\ref{mellinlog}). The first term in the r.h.s. of Eq.~(\ref{eqfmv}) 
corresponds to graphs (d,e) in Fig.~1. It relates  $F_{ij} (\omega,y_1,y_2)$ to partly off-shell amplitudes $F^{\prime}_{ij}$. The last term in  the r.h.s. of Eq.~(\ref{eqfmv}) 
corresponds to the sum of graphs    
(b) and (c) in Fig.~1 as well as the mirror graphs. It involves the color octet amplitude (the blobs on graphs (b,c) and mirror graphs), so we provide it with the 
superscript $(8)$. Neither Born term nor graph (a) in Fig.~1 contribute to Eq.~(\ref{eqfmv}): the Born term 
does not depend on $\mu$ when the external partons are virtual, so it vanishes when differentiated over $\mu$;
 graph (a) does not yield DL contribution at $t = 0$. So, Eq.~(\ref{eqfmv})  makes it possible to express 
 $F_{ij} (\omega,y_1,y_2)$ through amplitudes $F^{\prime}_{ij}$ which are simpler than $F_{ij} (\omega,y_1,y_2)$. 
 We compose and solve IREEs for them in the next Sect. 
 However, it involves also the octet contributions $\widetilde{T}_{ij} (\omega,y_1,y_2)$ which are quite complicated. 
 IREEs for them involve partly off-shell octet amplitudes, so these IREEs should be constructed and solved before working on Eq.~(\ref{eqfmv}) 
 so that their solutions could be substituted in Eq.~(\ref{eqfmv}).  This problem has not been solved up to now, although 
  our work on it is underway, 
 so we show in Sect.~VIII how to approximately account for $\widetilde{T}_{ij}$. \\
 The main difference between Eq.~(\ref{eqfmv}) and IREEs for the perturbative components of the structure function $g_1$ is that 
the latter do not involve  $\widetilde{T}_{ij} (\omega,y_1,y_2)$ because in this case the upper partons in Fig.~1 are replaced by 
virtual photons which cannot be coupled to the lower partons.  
It is shown in Sect.~VIII that this difference does not affect the small-$x$ behavior of the parton helicities. 

 Eq.~(\ref{eqfmv}) demonstrates  
the general strategy of the IREE method: IR equations with $n$ arguments involve simpler amplitudes 
depending on $n-1$ arguments; then, IREEs for them involve 
amplitudes depending on $n-2$ arguments and so on until arriving at amplitudes depending on one variable only. 
Now we proceed to solving Eq.~(\ref{eqfmv}). It looks simpler in terms of variables $z_{1,2}$ defined as follows: 
 
\begin{eqnarray}\label{yz}
z_1 = y_1 + y_2, 
\\ \nonumber
z_2 = y_1 - y_2. 
\end{eqnarray}

  The inverse relations are 

\begin{eqnarray}\label{invyz}
y_1 = (z_1 + z_2)/2, 
\\ \nonumber
y_2 = (z_1 - z_2)/2. 
\end{eqnarray}

In terms of $z_{1,2}$, Eq.~(\ref{eqfmv}) is

\begin{eqnarray}\label{eqfzoffin}
\left[\frac{\partial }{\partial z_1}  
+ \frac{ \omega}{2} \right] F_{ij} (\omega,y_1,y_2)   &=& 
\frac{1}{16 \pi^2} 
F^{\prime}_{ir} (\omega, y_1) F^{\prime}_{rj} (\omega, y_2) + \frac{1}{16 \pi^2} \widetilde{T}_{ij} (\omega,y_1,y_2).
\end{eqnarray}

In order to simplify Eq.~(\ref{eqfzoffin}), introduce auxiliary amplitudes $\Psi_{ij},\psi_{ij}$ and $V^{(8)}_{ij}$: 

\begin{eqnarray}\label{psivdef}
\Psi_{ij} = \frac{1}{8 \pi^2} F_{ij}, 
\\ \nonumber
\psi_{rj} = \frac{}{8 \pi^2} F^{\prime}_{rj}, 
\\  \nonumber
\widetilde{V}_{ij} = \frac{}{8 \pi^2} \widetilde{T}_{ij}. 
\end{eqnarray}

Then obtain from (\ref{eqfzoffin})

\begin{eqnarray}\label{eqpsiz}
\left[\frac{\partial}{\partial z_1}  + \frac{\omega}{2}\right] \Psi_{ij} (\omega,y_1,y_2)  
    &=&  \frac{1}{2}\left[
\psi_{ir} (\omega, y_1) \psi_{rj} (\omega, y_2) + \widetilde{V}_{ij} (\omega,y_1,y_2)
\right].
\end{eqnarray}

The general solution to it is:

\begin{eqnarray}\label{gensolpsic}
\Psi_{ij} (\omega,y_1,y_2)   &=& e^{-\omega z_1/2}\;\widetilde{C}_{ij} (\omega,z_1,z_2),
\end{eqnarray} 
with $\widetilde{C}_{ij}$ obeying 

\begin{eqnarray}\label{gensolwc}
\frac{\partial \widetilde{C}_{ij}}{\partial z_1}   &=& \frac{e^{\omega z_1/2}}{2}\;
\left[
\psi_{ir} (\omega, y_1) \psi_{rj} (\omega, y_2) + \widetilde{V}_{ij} (\omega,y_1,y_2)
\right].
\end{eqnarray} 

General solution to (\ref{gensolwc}) is 

\begin{eqnarray}\label{gensolc}
\widetilde{C}_{ij}   &=& \frac{1}{2} \int_0^{z_1} d z^{\prime}_1
e^{\omega z^{\prime}_1/2}\;
\left[
\psi_{ir} (\omega, y^{\prime}_1) \psi_{rj} (\omega, y^{\prime}_2) + \widetilde{V}_{ij} (\omega,y^{\prime}_1,y^{\prime}_2)
\right] + C (\omega, z_2),
\end{eqnarray} 
where $C (\omega, z_2)$ is arbitrary while $y^{\prime}_{1,2}$ are defined as follows: 

\begin{eqnarray}\label{yprime}
y^{\prime}_1 = (z^{\prime}_1 + z_2)/2, 
\\ \nonumber
y^{\prime}_2 = (z^{\prime}_1 - z_2)/2. 
\end{eqnarray}

In order to specify the general solution we first assume that 

\begin{equation}\label{y1y2ord}
y_1 > y_2
\end{equation}
and then use matching: 

\begin{equation}\label{matchfmv}
F_{ij} (\omega,y_1,y_2)|_{y_2 = 0}= F^{\prime}_{ij} (\omega,y_1),
\end{equation}
i.e. 

\begin{equation}\label{matchpsi}
\Psi_{ij} (\omega,y_1,y_2)|_{y_2 = 0}= \psi_{ij} (\omega,y_1),
\end{equation}
where $F^{\prime}_{ij} (\omega,y_1)$ denotes partly off-hell amplitudes and $\psi_{ij} (\omega,y_1)$ 
is defined in Eq.~(\ref{psivdef}). Obtain 

\begin{eqnarray}\label{psimv}
\Psi_{ij} (\omega,y_1,y_2)   &=& e^{-\omega y_2} \psi_{ij} (\omega,z_2) 
\\ \nonumber
&+& 
 \frac{e^{-\omega z_1/2} }{2} \int_{z_2}^{z_1} d z^{\prime}_1 e^{\omega z^{\prime}_1/2} 
\left[
\psi_{ir} (\omega, y^{\prime}_1) \psi_{rj} (\omega, y^{\prime}_2) + \widetilde{V}_{ij} (\omega,y^{\prime}_1,y^{\prime}_2)
\right] 
\end{eqnarray}
and therefore the amplitude $A_{ij}\left(w, Q^2,k^2_{\perp}\right)$ in the MV kinematics is 

 \begin{equation}\label{amv}
 A_{ij}\left(w, Q^2,k^2_{\perp}\right) = \frac{1}{8 \pi^2}\int_{-\imath \infty}^{\imath \infty} \frac{d \omega}{2 \pi \imath} 
 \left(w/\mu^2\right)^{ \omega}
  \Psi_{ij}\left(\omega, Q^2,k^2_{\perp}\right). 
 \end{equation}

It is easy to check that Eq.~(\ref{psimv}) satisfies both matching (\ref{matchpsi}) and (\ref{eqpsiz}).  
 It expresses the totally off-shell amplitudes through less complicated partly off-shell amplitudes. The ordering (\ref{y1y2ord}) can easily be 
 lifted with replacement $y_2$ by $|y_2|$. 

\subsection{Totally off-shell amplitudes in Deeply-Virtual kinematics}
In contrast to 
$A_{ij}(w,Q^2, k^2_{\perp})$, amplitudes $A^{DV}_{ij}(w,Q^2, k^2_{\perp})$ do not depend on $\mu$, i.e. they are infrared-stable, so IREEs for them are much simpler: 

\begin{equation}\label{eqadv}
\left[\frac{\partial}{\partial \rho} + \frac{\partial}{\partial y_1} + \frac{\partial}{\partial y_2} \right] A^{DV}_{ij}(w,Q^2, k^2_{\perp}) = 0
\end{equation}  
and the general solution to Eq.~(\ref{eqadv}) is also simple: 

\begin{equation}\label{gensoladv}
A^{DV}_{ij}(\rho,y_1,y_2) = C^{DV}_{ij} (\rho - y_1, \rho - y_2),
\end{equation}
where $C^{DV}_{ij}$ are arbitrary analytic  functions.
 We specify it by matching $A^{DV}_{ij}$  with amplitude $A_{ij}$ of  
 Eq.~(\ref{amv}) on the border between MV and DV kinematics, where 

\begin{equation}\label{border}
\rho = y_1 + y_2.
\end{equation}

The matching yields

\begin{equation}\label{matchamvdv}
C^{DV}_{ij} (\rho - y_1, \rho_ - y_2)|_{\rho = y_1 + y_2} = A_{ij} (\rho, y_1,y_2)|_{\rho = y_1 + y_2} 
\end{equation}
and therefore 

\begin{equation}\label{matchborder}
C^{DV}_{ij} (y_2, y_1)  = A_{ij} (y_1 + y_2, y_1,y_2),  
\end{equation}  
which fixes $C^{DV}_{ij}$ on the border (\ref{border}). In the region of DV kinematics (\ref{dvkin},\ref{dvkinx}), 
i.e. off the border, 
$C^{DV}_{ij} (\rho - y_1, \rho_ - y_2)$ is 
obtained out of $A_{ij} (y_1 + y_2, y_1,y_2) $ with replacements $y_1 \to \rho - y_2, y_2 \to \rho - y_1$, 
i.e. $z_1 \to 2 \rho - y_1 - y_2 = \rho - z_1$ and $z_2 \to y_2 - y_1 = - z_2$: 

\begin{equation}\label{advamv}
A^{DV}_{ij} (\rho,y_1, y_2) = A_{ij} (2\rho- y_1 - y_2,\rho - y_2, \rho - y_1).  
\end{equation}
and 

\begin{equation}\label{mdvmmv}
M^{DV}_{ij} (\rho,y_1, y_2) = M_{ij} (2\rho- y_1 - y_2,\rho - y_2, \rho - y_1).  
\end{equation}

The specific form of dependence of the r.h.s. of Eqs.~(\ref{advamv}, \ref{mdvmmv}) on $\rho, y_{1,2}$ clearly demonstrates that $A^{DV}_{ij}$ 
and $M^{DV}_{ij}$ do not depend on the IR 
cut-off, i.e. these expression are IR stable. 

\section{Partly off-shell  amplitudes} 

Being guided by the general prescription (see Sect.~IIC) for composing IREEs and remembering how Eq.~(\ref{eqfmv}) was constructed, one can easily 
compose IREEs for  
$F^{\prime}_{ij}(\omega,y)$ or related to them amplitudes $\psi_{ij}$ (see Eq.~(\ref{psivdef})).  Actually, the system of IREEs for $\psi_{ij}$ splits into two subsystems:

\begin{eqnarray}\label{eqpsi1}
\frac{\partial \psi_{qq}\left(\omega, y\right)}{\partial y} &=&
\psi_{qq}\left(\omega, y\right)h_{qq} (\omega) + \psi_{qg}\left(\omega, y\right)h_{gq} (\omega) + V^{\prime}_{qq}\left(\omega, y\right),
\\ \nonumber
\frac{\partial \psi_{qg}\left(\omega, y\right)}{\partial y} &=& \psi_{qq}\left(\omega, y\right)h_{qg} (\omega) +
\psi_{qg}\left(\omega, y\right)h_{gg} (\omega) + V^{\prime}_{gq}\left(\omega, y\right)
\end{eqnarray}
and

\begin{eqnarray}\label{eqpsi2}
\frac{\partial \psi_{gq}\left(\omega, y\right)}{\partial y} &=&
\psi_{gq}\left(\omega, y\right)h_{qq} (\omega) + \psi_{gg}\left(\omega, y\right)h_{gq} (\omega) + V^{\prime}_{qg}\left(\omega, y\right),
\\ \nonumber
\frac{\partial \psi_{gg}\left(\omega, y\right)}{\partial y} &=& \psi_{gq}\left(\omega, y\right)h_{qg} (\omega) +
\psi_{gg}\left(\omega, y\right)h_{gg} (\omega) + V^{\prime}_{gg}\left(\omega, y\right),
\end{eqnarray}
where $ V^{\prime}_{ij}\left(\omega, y\right)$ stand for color octets. They correspond to the sum of convolutions   
(a, c) and mirror in Fig.~1. Contribution of the Born term vanishes when differentiated over $\mu$.  Amplitudes $h_{ij}$ 
do not depend on $y$, so we refer to them as on-shell amplitudes. Let us note once 
more that Eq.~(\ref{eqfmv}) differs from IREEs for the DIS structure function $g_1$ in that the latter do not involve 
octet contributions.

First step to solve Eqs.~(\ref{eqpsi1},\ref{eqpsi2}) is to find general solutions to homogeneous equations, where $V^{\prime}_{ij} = 0$: 

 \begin{eqnarray}\label{gensolpsi1}
\psi_{qq} &=& C_1(\omega) e^{y \Omega_{+}} + C_2(\omega) e^{y \Omega_{-}},
\\ \nonumber
\psi_{qg} &=& C_1(\omega)r_1 (\omega) e^{y \Omega_{(+)}} + C_2(\omega)r_2 (\omega) e^{y \Omega_{(-)}},
\end{eqnarray}

and 

\begin{eqnarray}\label{gensolpsi2}
\psi_{gq} &=& C_3(\omega) r_3 (\omega) e^{y \Omega_{(+)}} + C_4(\omega) r_4 (\omega) e^{y \Omega_{(-)}},
\\ \nonumber
\psi_{gg} &=& C_3 (\omega)e^{y \Omega_{+}} + C_4(\omega)e^{y \Omega_{-}},
\end{eqnarray}
where 
\begin{equation}\label{omegapm}
\Omega_{(\pm)} = \frac{1}{2} \left[h_{qq} + h_{gg} \pm \sqrt{R}\right],
\end{equation}
with

\begin{equation}\label{r}
R = (h_{gg} - h_{qq})^2 + 4 h_{qg}h_{gq}
\end{equation}
and
\begin{eqnarray}\label{bj}
r_1 &=& 
\frac{1}{2h_{gq}} \left[h_{gg} - h_{qq} + \sqrt{R}\right],
\\ \nonumber
r_2 &=& 
\frac{1}{2h_{gq}} \left[h_{gg} - h_{qq} - \sqrt{R}\right],
\\ \nonumber
r_3 &=& 
\frac{1}{2h_{qg}} \left[h_{qq} - h_{gg} - \sqrt{R}\right],
\\ \nonumber
r_4 &=& 
\frac{1}{2h_{qg}} \left[h_{qg} - h_{gg} + \sqrt{R}\right].
\end{eqnarray}
The factors $C_{1,2,3,4} (\omega)$ in Eqs.~(\ref{gensolpsi1},\ref{gensolpsi2}) are arbitrary. Let us focus on Eq.~(\ref{gensolpsi1}). It can be solved with applying standard 
mathematical means. 
Put $C_{1,2}$ be dependent on $y$ and substitute it in (\ref{eqpsi1}):

\begin{eqnarray}\label{eqcprime}
d C_1/dy\; e^{y \Omega_{+}} + dC_2/dy\; e^{y \Omega_{-}} &=&  V^{\prime}_{qq}\left(\omega, y\right),
\\ \nonumber
d C_1/dy\;r_1\; e^{y \Omega_{+}} + d C_2/dy\; r_2\;e^{y \Omega_{-}} &=&  V^{\prime}_{qg}\left(\omega, y\right).
\end{eqnarray}

 Solution to (\ref{eqcprime}) is:

\begin{eqnarray}\label{solcprime}
\frac{dC_1\left(\omega, y\right)}{dy} &=&  e^{- y \Omega_{+}}~ \frac{V^{\prime}_{qq}\left(\omega, y\right)- r_2\;V^{\prime}_{qg} \left(\omega, y\right)}{r_2 - r_1},
\\ \nonumber
\frac{d C_2\left(\omega, y\right)}{dy}  &=&  e^{- y \Omega_{-}}~ \frac{ - r_1\;V^{\prime}_{qq}\left(\omega, y\right)  + V^{\prime}_{qg} \left(\omega, y\right)}{r_2 - r_1} .
\end{eqnarray}

Integrating (\ref{solcprime}) yields:

\begin{eqnarray}\label{solc}
C_1\left(\omega, y\right) &=& \widetilde{C}_1 (\omega)  + \frac{1}{r_2 - r_1}\int_0^y d ^{\prime}  e^{- y^{\prime} \Omega_{+}} 
\left[V^{\prime}_{qq}\left(\omega, y^{\prime}\right)- r_2\;V^{\prime}_{qg} \left(\omega, y^{\prime}\right)\right],
\\ \nonumber
C_2\left(\omega, y\right) &=& \widetilde{C}_2 (\omega)  + \frac{1}{r_2 - r_1}\int_0^y d ^{\prime}  e^{- y^{\prime} \Omega_{-}} 
\left[- r_1\;V^{\prime}_{qq}\left(\omega, y\right)  + V^{\prime}_{qg} \left(\omega, y\right)\right], 
\end{eqnarray}
where $\widetilde{C}_{1,2}$ do not depend on $y$. Substituting it in (\ref{gensolpsi1}), obtain 

 \begin{eqnarray}\label{solpsi1}
\psi_{qq} &=&  e^{y \Omega_{+}} \left[\widetilde{C}_1 (\omega)  + \frac{1}{r_2 - r_1}\int_0^y d ^{\prime}  e^{- y^{\prime} \Omega_{+}} 
\left[V^{\prime}_{qq}\left(\omega, y^{\prime}\right)- r_2\;V^{\prime}_{qg} \left(\omega, y^{\prime}\right)\right]\right]
\\ \nonumber
&+&  e^{y \Omega_{-}} 
\left[\widetilde{C}_2 (\omega)  + \frac{1}{r_2 - r_1}\int_0^y d ^{\prime}  e^{- y^{\prime} \Omega_{-}} 
\left[- r_1\;V^{\prime}_{qq}\left(\omega, y\right)  + V^{\prime}_{qg} \left(\omega, y\right)\right]\right],
\\ \nonumber
\psi_{qg} &=&  r_1\;e^{y \Omega_{+}} \left[\widetilde{C}_1 (\omega)  + \frac{1}{r_2 - r_1}\int_0^y d ^{\prime}  e^{- y^{\prime} \Omega_{+}} 
\left[V^{\prime}_{qq}\left(\omega, y^{\prime}\right)- r_2\;V^{\prime}_{qg} \left(\omega, y^{\prime}\right)\right]\right]
\\ \nonumber
&+&  r_2\;e^{y \Omega_{-}} 
\left[\widetilde{C}_2 (\omega)  + \frac{1}{r_2 - r_1}\int_0^y d ^{\prime}  e^{- y^{\prime} \Omega_{-}} 
\left[- r_1\;V^{\prime}_{qq}\left(\omega, y\right)  + V^{\prime}_{qg} \left(\omega, y\right)\right]\right]. 
\end{eqnarray}

In order to fix $\widetilde{C}_{1,2} (\omega)$, we use the matching with the on-shell amplitudes $h_{ij} (\omega)$: 

\begin{equation}\label{matchpsih1}
\psi_{qq} (\omega,y)|_{y = 0} = h_{qq} (\omega),~\psi_{qg} (\omega,y)|_{y = 0} = h_{qg} (\omega).
\end{equation}

It yields 

\begin{eqnarray}\label{solc1}
\widetilde{C}_1 &=& \frac{r_2 h_{qq} - h_{qg}}{r_2 - r_1},
\\ \nonumber
\widetilde{C}_2 &=& \frac{-r_1 h_{qq} + h_{qg}}{r_2 - r_1}
\end{eqnarray}
and therefore (\ref{solpsi1}) takes the following form: 

 \begin{eqnarray}\label{solpsi1h}
\psi_{qq} &=&  \frac{e^{y \Omega_{+}}}{r_2 - r_1} \left[ \left(r_2 h_{qq} - h_{qg}\right)  +\int_0^y d ^{\prime}  e^{- y^{\prime} \Omega_{+}} 
\left[V^{\prime}_{qq}\left(\omega, y^{\prime}\right)- r_2\;V^{\prime}_{qg} \left(\omega, y^{\prime}\right)\right]\right]
\\ \nonumber
&+&  \frac{e^{y \Omega_{-}}}{r_2 - r_1} 
\left[\left(-r_1 h_{qq} + h_{qg}\right)  + \int_0^y d ^{\prime}  e^{- y^{\prime} \Omega_{-}} 
\left[- r_1\;V^{\prime}_{qq}\left(\omega, y\right)  + V^{\prime}_{qg} \left(\omega, y\right)\right]\right],
\\ \nonumber
\psi_{qg} &=&  \frac{r_1\;e^{y \Omega_{+}}}{r_2 - r_1} \left[ \left(r_2 h_{qq} - h_{qg}\right) + \int_0^y d ^{\prime}  e^{- y^{\prime} \Omega_{+}} 
\left[V^{\prime}_{qq}\left(\omega, y^{\prime}\right)- r_2\;V^{\prime}_{qg} \left(\omega, y^{\prime}\right)\right]\right]
\\ \nonumber
&+&  \frac{r_2\;e^{y \Omega_{-}}}{r_2 - r_1} 
\left[ \left(-r_1 h_{qq} + h_{qg}\right) + \int_0^y d ^{\prime}  e^{- y^{\prime} \Omega_{-}} 
\left[- r_1\;V^{\prime}_{qq}\left(\omega, y\right)  + V^{\prime}_{qg} \left(\omega, y\right)\right]\right]. 
\end{eqnarray}

Expressions (\ref{eqpsi2}) for $\psi_{gq}$ and $\psi_{gg}$ can be solved in the same way, however with using the 
other matching: 

\begin{equation}\label{matchpsih2}
\psi_{gq} (\omega,y)|_{y = 0} = h_{gq} (\omega),~\psi_{gg} (\omega,y)|_{y = 0} = h_{gg} (\omega). 
\end{equation}

As a result, obtain  

 \begin{eqnarray}\label{solpsi2h}
\psi_{gg} &=&  \frac{e^{y \Omega_{+}}}{r_4 - r_3} \left[ \left(r_4 h_{gg} - h_{gq}\right)  +\int_0^y d ^{\prime}  e^{- y^{\prime} \Omega_{+}} 
\left[V^{\prime}_{gg}\left(\omega, y^{\prime}\right)- r_4\;V^{\prime}_{gq} \left(\omega, y^{\prime}\right)\right]\right]
\\ \nonumber
&+&  \frac{e^{y \Omega_{-}}}{r_4 - r_3} 
\left[\left(-r_3 h_{gg} + h_{gq}\right)  + \int_0^y d ^{\prime}  e^{- y^{\prime} \Omega_{-}} 
\left[- r_1\;V^{\prime}_{gg}\left(\omega, y\right)  + V^{\prime}_{gq} \left(\omega, y\right)\right]\right],
\\ \nonumber
\psi_{gq} &=&  \frac{r_1\;e^{y \Omega_{+}}}{r_4 - r_3} \left[ \left(r_4 h_{gg} - h_{gq}\right) + \int_0^y d ^{\prime}  e^{- y^{\prime} \Omega_{+}} 
\left[V^{\prime}_{gg}\left(\omega, y^{\prime}\right)- r_4\;V^{\prime}_{gq} \left(\omega, y^{\prime}\right)\right]\right]
\\ \nonumber
&+&  \frac{r_4\;e^{y \Omega_{-}}}{r_4 - r_3} 
\left[ \left(-r_4 h_{gg} + h_{gq}\right) + \int_0^y d ^{\prime}  e^{- y^{\prime} \Omega_{-}} 
\left[- r_3\;V^{\prime}_{qq}\left(\omega, y\right)  + V^{\prime}_{gq} \left(\omega, y\right)\right]\right]. 
\end{eqnarray}

Eqs.~(\ref{solpsi1h},\ref{solpsi2h}) represent the partly off-shell amplitudes in terms of on-sell amplitudes $h_{ij}(\omega)$ and the color octet 
contributions $V^{\prime}_{ij}$. First, we proceed to calculating the on-shell amplitudes.

\section{On-shell amplitudes} 

IREEs for the on-shell amplitudes $f_{ij} (\omega)$ can also be obtained from the general pattern presented in 
Sect.~IIC and  depicted in 
Fig.~1 in the same way as the IREEs for the off-shell amplitudes considered above.  
However, they differ from  off-shell equations. First, the on-shell l.h.ss. do not involve 
derivatives because $f_{ij} (\omega)$ do not depend on $y$, which leads to algebraic equations. Second, the 
on-shell Born terms  
depend on $\mu$, so, in contrast to Eqs.~(\ref{eqfmv},\ref{eqpsi1},\ref{eqpsi2}), they contribute to  
on-shell IREEs.   

\subsection{IREEs for on-shell amplitudes}

Accounting for all terms in Fig.~1 leads to the following IREEs: 

\begin{equation}\label{eqfon}
\omega f_{ij}(\omega) = a^B_{ij} (\omega) + \frac{1}{8 \pi^2}\; f_{in}(\omega) f_{nj}(\omega) + \frac{1}{8 \pi^2}\;t^{(8)}_{ij}(\omega),
\end{equation}
where $ a^B_{ij}$ stand for the Born terms, $n = q,g$ and $t^{(8)}(\omega)$ corresponds to the sum of convolutions (b,c) in Fig.~1 when all external partons are on-shell. 
The contribution of graph (a) does not yield double-logs because $t = 0$, so this graph is dropped. 
As in the previous cases, we introduce amplitudes $h_{ij}$ instead of $f_{ij}$: 

\begin{equation}\label{fh}
h_{ij}(\omega) = \frac{1}{8 \pi^2}\; f_{ij}(\omega). 
\end{equation}
  
 In terms of $h_{ij}$, IREEs for the on-shell amplitudes are as follows: 

\begin{eqnarray}\label{eqhon} 
 \omega  h_{qq} (\omega)   &=& b_{qq} + 
h_{qq} (\omega) h_{qq} (\omega) + 
 h_{qg} (\omega) h_{gq} (\omega), 
\\   \nonumber  
 \omega h_{qg} (\omega)   &=& b_{qg} + 
h_{qq} (\omega) h_{qg} (\omega) + 
 h_{qg} (\omega) h_{gg} (\omega), 
\\   \nonumber
 \omega  h_{gq} (\omega)   &=&  b_{gq} + 
b_{gq} (\omega) h_{qq} (\omega) + 
 h_{gg} (\omega) h_{gq} (\omega), 
\\   \nonumber 
\omega  h_{gg} (\omega)   &=&  b_{gg} + 
h_{gq} (\omega) h_{qg} (\omega) + 
 h_{gg} (\omega) h_{gg} (\omega), 
\end{eqnarray} 
where inhomogeneous terms $b_{ij}$ contain the Born contributions and contributions of the color octets, see Ref.~\cite{egtg1s}: 

\begin{equation}\label{bav}
b_{ij}(\omega) = a_{ij}(\omega) + V_{ij}(\omega), 
\end{equation}
with 

\begin{equation}\label{av}
a_{ij}(\omega) = \frac{1}{8 \pi^2}\; a^B_{ij}(\omega),~ V_{ij}(\omega) = \frac{1}{8 \pi^2}\;t^{(8)}_{ij}(\omega).
\end{equation}
 
The Born contributions $a_{ij}$ can be found in Ref.~\cite{egtg1s,egtg1sum} and discussion on the color octets is in Sect.~VIII.  

\subsection{Solutions to the equations for the on-shell-amplitudes}

Equations in Eq.~(\ref{eqhon}) for the on-shell amplitudes $h_{ij}$ are algebraic but  non-linear. 
Solution to them can be found 
analytically. We remind that the DIS structure function $g_1$ in DLA is also made out of amplitudes $h_{ij}$.  
Approximate solution to Eq.~(\ref{eqhon}) in the analytic form was presented in Refs.~\cite{egtg1s, egtg1sum}. 
Below we 
represent this solution in a more convenient form.
First, introduce auxiliary amplitudes 

\begin{equation}\label{hpm}
h_{\pm} = h_{qq} \pm h_{gg}
\end{equation}
and express $h_{qg}$ and $h_{gg}$ in terms of  $h_{+}$: 

\begin{equation}\label{hqghplus}
\omega h_{qg} = \frac{b_{qg}}{\omega - h_+},~~h_{gq} = \frac{b_{gq}}{\omega - h_+},
\end{equation}

Then rewrite the remaining equations in (\ref{eqhon}) in terms of $h_{\pm}$: 

\begin{eqnarray}\label{hphm}
h_{-} &=& \frac{b_{-}}{\omega - h_+},
\\ \nonumber
\omega h_+ &=& b_{+} + h^2_{qq} + h^2_{gg} + 2b_{qg} b_{gq}/(\omega - h_+)^2,
\end{eqnarray}
where 

\begin{equation}\label{bpm}
b_{\pm} = b_{qq} \pm b_{gg}. 
\end{equation}

Replace $h^2_{qq} + h^2_{gg}$ by $(1/2) \left[h^2_+ + h^2_- \right]$ and arrive at the following equation for $h_+$: 

\begin{eqnarray}\label{eqhplus}
\omega h_+ &=& b_{+} + \frac{1}{2} \left[h^2_+ + \frac{(b_{qq} - b_{gg})^2}{(\omega - h_+)^2}\right]
 + \frac{2b_{qg} b_{gq}}{(\omega - h_+)^2}
 \\ \nonumber
&=& b_{+} + \frac{1}{2} \left[(\omega -h_+)^2 + 2 \omega h_+ - \omega^2  + \frac{(b_{qq} - b_{gg})^2}{(\omega - h_+)^2}\right]
 + \frac{2b_{qg} b_{gq}}{(\omega - h_+)^2}.
\end{eqnarray}

Denoting $(\omega - h_+)^2 = u$, obtain an algebraic quadratic equation for $u$: 

\begin{equation}\label{equ}
u^2 -[ \omega^2 -2 (b_{qq} + b_{gg})]\;u + [(b_{qq} - b_{gg})^2 + 4b_{qg} b_{gq}]  = 0. 
\end{equation}

This equation has four roots but only one of them satisfies matching with the Born amplitudes 

\begin{equation}\label{matchborn}
h_{ik} = \frac{b_{ik}}{\omega}
\end{equation}
at $\omega \gg 1$. As a result we obtain explicit expressions for $h_{ik}$: 

\begin{eqnarray}\label{hz}
&& h_{qq} = \frac{1}{2} \Big[ \omega - Z - \frac{b_{gg} -
b_{qq}}{Z}\Big],\qquad h_{qg} = \frac{b_{qg}}{Z}~, \\ \nonumber &&
h_{gg} = \frac{1}{2} \Big[ \omega - Z + \frac{b_{gg} -
b_{qq}}{Z}\Big],\qquad h_{gq} =\frac{b_{gq}}{Z}~,
\end{eqnarray}
where

\begin{equation}\label{zuw}
Z = \sqrt{ \frac{1}{2} \left[ U + \sqrt{W} \right]},
\end{equation}
with 

\begin{equation}\label{u}
U = \omega^2 - 2b_{+}
\end{equation}
and

\begin{equation}\label{w}
W = U^2  -4 (b^2_{-} + 4b_{qg} b_{gq}) =(\omega^2 - 2b_{+})^2 -4 (b^2_{-} + 4b_{qg} b_{gq}).
\end{equation}

Making use of Eqs.~(\ref{zuw} \ref{w}) brings (\ref{hz}) to the more convenient form: 

\begin{eqnarray}\label{huw}
h_{qq} &=&  \frac{1}{2}\left[\omega - \frac{\sqrt{U + \sqrt{W}}}{{\sqrt{2}}} + 
\frac{b_{-}}{\sqrt{b^2_{-} + 4 b_{qg}b_{gq}}} \frac{\sqrt{U - \sqrt{W}}}{{\sqrt{2}}}\right],
\\ \nonumber
h_{gg} &=&  \frac{1}{2}\left[\omega - \frac{\sqrt{U + \sqrt{W}}}{{\sqrt{2}}} - 
\frac{b_{-}}{\sqrt{b^2_{-} + 4 b_{qg}b_{gq}}} \frac{\sqrt{U - \sqrt{W}}}{{\sqrt{2}}}\right],
\\ \nonumber
h_{qg} &=& \frac{b_{qg}}{\sqrt{b^2_{-} + 4 b_{qg}b_{gq}}} \frac{\sqrt{U - \sqrt{W}}}{{\sqrt{2}}},
\\ \nonumber
h_{gq} &=& \frac{b_{gq}}{\sqrt{b^2_{-} + 4 b_{qg}b_{gq}}} \frac{\sqrt{U - \sqrt{W}}}{{\sqrt{2}}}.
\end{eqnarray}

\section{Remark on octet contributions}

Terms $\widetilde{V}_{ij}, V^{\prime}_{ij}$ and $V_{ij}$ in Eqs.~(\ref{eqpsiz},\ref{eqpsi1},\ref{eqpsi2}) and (\ref{eqhon}) 
involve color octet amplitudes in the following way: 

\begin{equation}\label{tv}
\hat{V}_{ij} = \kappa_{ij} \frac{\alpha_s}{\pi} \frac{\hat{F}^{(8)}}{\omega^2},
\end{equation}
where $\kappa_{ij}$ are the color factors while $\hat{V}_{ij}$ and $\hat{F}^{(8)}$  
 are generic notations of color octet amplitudes 
and their contributions to IREEs  in the kinematics considered above (totally off-shell, partly off-shell  and on-shell kinematics). 
IREEs for $\hat{F}^{(8)}$ are more involved than ones for the color singlets. Strictly speaking, only one particular 
case has been studied so far.  
Namely, there was obtained and solved the IREE for the on-shell amplitude $h^{(8)}_{qq} (\omega)$ whereas  
all other $h^{(8)}_{ij} (\omega)$ were neglected (flavour non-singlet case).  This equation was obtained in Ref.~\cite{kl1,kl2}:  

\begin{equation}\label{eqh8qq}
\omega h^{(8)}_{qq} (\omega) = a^{(8)}_{qq} + \kappa_{qq}\frac{\alpha_s}{\pi}
\frac{d}{d \omega} \;h^{(8)}_{qq} (\omega) +  
h^{(8)}_{qq} (\omega) h^{(8)}_{qq} (\omega),
\end{equation}
with $a^{(8)}_{qq} = - \alpha_s/(4 \pi N)$ and $\kappa_{qq} = 2 C_F$. Solution to this equation is 

\begin{equation}\label{h8qq}
h^{(8)}_{qq} (\omega) = \frac{\alpha_s N}{2 \pi} \frac{d}{d \omega} \ln \left[e^{z^2/4} D_p (z)\right],
\end{equation}
where $D_p(z)$ denotes the Parabolic Cylinder Function, with $z = \omega/\sqrt{\alpha_s N/(2 \pi)}$ and $p = - 1/(2 N^2)$. 
Straightforward substitution of  Eq.~(\ref{h8qq}) in 

\begin{equation}\label{hqq}
h_{qq} = \frac{1}{2} \sqrt{\omega^2 - 4 b_{qq}(\omega)},
\end{equation}
where $b_{qq}$ is given by (\ref{bav}), with $a_{qq} = \alpha_s C_F/(2 \pi)$ and $\kappa_{qq} = N/2$, leads to the 
very complicated expression which can be treated with numerical calculations only. However, $h^{(8)}_{qq} (\omega)$ 
quickly decreases at large $\omega$. This makes it possible to approximate $h^{(8)}_{qq} (\omega)$ by its Born value $a_{qq}/\omega$. 
Such an approximation was made in Refs.~\cite{kl1,kl2,berns} for color non-singlets and then generalized in Refs.~\cite{egtg1s, egtg1sum} 
to the case of the singlet $g_1$. Approximation expressions for $b_{ij}$ can be found in Refs.~\cite{egtg1s,egtg1sum}. 
Explicit expressions for the octet amplitudes contributing to Eqs.~(\ref{eqfmv}, \ref{eqpsi1},\ref{eqpsi2}) are unknown yet although 
our work on them is underway. So, in the present paper we account for them only approximately, through $V_{ij}$ in the Born approximation. 

It is worth mentioning in this regard that the impact of the octets on the intercepts of $h_{ij}$ and $g_1$ is known to be 
not so large. 
For instance, the 
intercept $\omega_0$ of the small-$x$ asymptotics of $h_{qq}$ is 

\begin{eqnarray}\label{omegaqq}
\omega_0 &=& \omega_0^{\prime} 
\sqrt{\frac{1 + \sqrt{1 + 2/(N C_F)}}{2}}
\approx  \omega_0^{\prime} 
\left[1 + 1/2N^2\right],
\end{eqnarray}
where $\omega_0^{\prime} = \sqrt{2 \alpha_s C_F/\pi}$ accommodates contributions of the ladder graphs 
only\footnote{Throughout the paper we imply the Feynman gauge for virtual gluons.}. The octet contribution in Eq.~(\ref{omegaqq}) is 
presented by the term $2/(N C_F)$ which stems from accounting for non-ladder graphs. 

\section{Expressions for helicities in the region $D_B$}

Expressions for the parton helicities are given in Eqs.~(\ref{factcol},\ref{factkt}) in the symbolic form. In the 
present Sect. we 
obtain explicit expressions for them in both Collinear and KT forms of QCD factorization. 

\subsection{Helicities in Collinear Factorization}

Collinear Factorization operates with partly on-shell amplitudes 
$M^{\prime}_{ij}$, where virtualities of the initial partons are fixed by the factorization scale $\mu^2$, so 

\begin{eqnarray}\label{hfactcol}
h_q \left(x, Q^2\right) &=& \int_{x}^1 d \beta \left[M^{\prime}_{qq}\left(x/\beta, Q^2\right)\otimes \Phi_q (\beta) + 
 M^{\prime}_{qg}\left(x/\beta, Q^2\right)\otimes \Phi_g (\beta)\right],
\\ \nonumber
h_g \left(x, Q^2\right) &=& \int_{x}^1 d \beta \left[ M^{\prime}_{gq}\left(x, Q^2\right)\otimes \Phi_q (x) +  
M^{\prime}_{gg}\left(x, Q^2\right)\otimes \Phi_g (x) \right].
\end{eqnarray}

Combining Eq.~(\ref{hfactcol}) and Eq.~(\ref{psivdef}), we obtain expressions 
for the parton helicities $h_{q,g}$ in the framework of Collinear Factorization: 

\begin{eqnarray}\label{hmelcol}
h_q \left(x, y_1\right) &=& \int_{-\imath \infty}^{\imath \infty} \frac{d \omega}{2 \pi \imath} x^{-\omega}
\;\omega\;
 \left[\psi_{qq} (\omega, y_1)\;\varphi_q (\omega) 
 + \psi_{qg} (\omega,y_1)\; \varphi_g (\omega)\right],
 \\ \nonumber
h_g \left(x, y_1\right) &=& \int_{-\imath \infty}^{\imath \infty} \frac{d \omega}{2 \pi \imath} x^{-\omega}
\;\omega\;
 \left[\psi_{gq} (\omega, y_1)\;\varphi_q (\omega) 
 + \psi_{gg} (\omega,y_1)\;\varphi_g (\omega)\right], 
 \end{eqnarray}
where $\psi_{ij}$ are given by Eqs.~(\ref{solpsi1h},\ref{solpsi2h}).  The parton distributions $\varphi_{q,g}$ will be specified 
in Sect.~XI. 

\subsection{Helicities in KT Factorization}

Expressions for $h_{q,g}$ in the framework of $KT$ Factorization in a generic form are given by  obtained by  
 Eq.~(\ref{factkt}.  We convert it in the constructive form, replacing the symbolic sign $\otimes$ 
 by explicit integrations over $\beta$ 
 and $k_{\perp}$. First step to do it is to fix the integration region which denote it $D_{KT}$. 
  The region $D_{KT}$ shown in Fig.~2. It is restricted by the condition $1 \geq \beta \geq x + k^2_{\perp}/w$ and 
  consists of  subregions $D_{MV}$ and $D_{DV}$. Each of them includes the condition $1 \geq \beta \geq x + k^2_{\perp}/w$ 
  and the restrictions for $k^2_{\perp}$ corresponding to Eqs.~(\ref{mvkin},\ref{dvkin}):

 \begin{equation}\label{mvirtx}
D_{MV}:~~ \mu^2/x \geq k^2_{\perp} \geq \mu^2,~1 \geq \beta \geq x + k^2_{\perp}/w,
 \end{equation}
 
\begin{equation}\label{dvirtx}
D_{DV}:~~  w (1-x) \geq k^2_{\perp} > \mu^2/x,~1 \geq \beta \geq x + k^2_{\perp}/w.
 \end{equation} 
 
So, we obtain 

\begin{eqnarray}\label{hkt}
h^{(KT)}_q \left(x, Q^2\right) &=& \int_{\mu^2}^{\mu^2/x} \frac{d k^2_{\perp}}{k^2_{\perp}}
\int_{x + k^2_{\perp}/w}^1  \frac{d \beta}{\beta} \left[M_{qq}\left(x/\beta, Q^2,k^2_{\perp}\right)\;\Phi^{KT}_q (\beta,k^2_{\perp}) + 
 M_{qg}\left(x/\beta, Q^2,k^2_{\perp}\right) \;\Phi^{KT}_g (\beta,k^2_{\perp})\right] 
 \\ \nonumber
&+& \int_{\mu^2/x}^{w (1-x)} \frac{d k^2_{\perp}}{k^2_{\perp}}
\int_{x+ k^2_{\perp}/w}^1  \frac{d \beta}{\beta} \left[ M^{DV}_{qq}\left(x/\beta, Q^2,k^2_{\perp}\right)\;\Phi^{KT}_q (\beta,k^2_{\perp}) +  
M^{DV}_{qg}\left(x/\beta, Q^2,k^2_{\perp}\right)\; \Phi^{KT}_g (\beta,k^2_{\perp}) \right],
\\ \nonumber
h^{(KT)}_g \left(x, Q^2\right) &=& \int_{\mu^2}^{\mu^2/x} \frac{d k^2_{\perp}}{k^2_{\perp}}
\int_{x + k^2_{\perp}/w}^1  \frac{d \beta}{\beta} \left[M_{gq}\left(x/\beta, Q^2,k^2_{\perp}\right)\;\Phi^{KT}_q (\beta,k^2_{\perp}) + 
 M_{gg}\left(x/\beta, Q^2,k^2_{\perp}\right) \;\Phi^{KT}_g (\beta,k^2_{\perp})\right]
 \\ \nonumber
&+& \int_{\mu^2/x}^{w (1-x)} \frac{d k^2_{\perp}}{k^2_{\perp}}
\int_{x+ k^2_{\perp}/w}^1  \frac{d \beta}{\beta} \left[ M^{DV}_{gq}\left(x/\beta, Q^2,k^2_{\perp}\right)\;\Phi^{KT}_q (\beta,k^2_{\perp}) +  
M^{DV}_{gg}\left(x/\beta, Q^2,k^2_{\perp}\right)\; \Phi^{KT}_g (\beta,k^2_{\perp}) \right],
\end{eqnarray}

with $M_{ij}$  defined in Eqs.~(\ref{psivdef},\ref{psimv}) and $M^{DV}_{ij}$ defined in Eq.~(\ref{mdvmmv}). 
Expressions for $h^{(KT)}_{q,g}$ in Eqs.~(\ref{hkt}) 
are more involved than the ones in Collinear Factorization but sometimes involving KT Factorization can be unavoidable. 
We consider one such situation below. 

\subsection{ $KT$ Factorization and  Orbital Angular Momenta}

$KT$ Factorization is more general form of QCD Factorization than Collinear one and  involves more 
complicated formulae than Collinear Factorization. 
It is the reason why Collinear Factorization is used more often than $KT$ Factorisation. However, there are situations when 
using $KT$ Factorization is mandatory.  In particular, it takes place when impact of   
 Orbital Angular Momenta (OAM) $L_{q,g}$ of the partons on the nucleon spin is investigated. Indeed, 
the conventional scenario of description of the nucleon spin $S_N$ is that $S_N$ is formed by spins of the partons (quarks and gluons) complemented 
by their OAMs $L_{q,g}$: 

\begin{equation}\label{ldef}
S_N = \frac{1}{2} \Delta \Sigma + \Delta G + L_q + L_g,
\end{equation}
where $(1/2)\Delta \Sigma$ and $\Delta G$ are conventional notations for contributions of the quark and gluon spins respectively. 
Aim of adding $L_{q,g}$ to them  in Refs.~\cite{oam1,oam2,oam3} was to increase the r.h.s. of Eq.~(\ref{ldef}). Extensive overview of the literature on 
OAM can be found in Ref.~\cite{leader}. 
Let us choose the frame where a high-energy nucleon moves along the $z$-axis, i.e. its momentum $\vec{P}$ 
has the only component $P_z$. Then the nucleon spin/helicity is also directed along the $z$-axis. 
Obviously, $L_{q,g}$ can increase the Total Angular Momentum if they have $z$-components $(L_{q,g})_z$.  
However, $(L_{q,g})_z \neq 0$ only when the parton momenta have transverse components. 
Indeed, consider a very simple example: the $z$-component of OAM $L_z$ of a parton with momentum $k$ is  

\begin{equation}\label{lz}
L_z = y k_x - x k_y,
\end{equation}
i.e. $\vec{k}$ has both longitudinal component $k_z$ and transverse component(s). The latter 
 excludes involvement of Collinear Factorization for evolving $L_{q,g}$
but leaves 
possible applying $KT$ Factorization. 
The above arguments hold when $L_{q,g}$ is applied to other 
hadronic reactions, see e.g. the recent paper\cite{hatta2}.  In order to avoid misunderstanding, we stress that we 
are not against using  
OAM contributions for description of longitudinal nucleon spin but we argue against using Collinear Factorization for this reason. 
We remind also that involvement of Collinear Factorization is not considered in Refs.~\cite{oam1,oam2,oam3,leader}.

\section{Helicities in Collinear Factorization at arbitrary $x$ and $Q^2$}

DGLAP evolution equations\cite{dglap1,dglap2,dglap3,dglap4} can be used for calculation of the parton helicities $h^{DGLAP}_{q,g}$ at large $x$ and $Q^2$, i.e. in region $D_A$. 
The DGLAP coefficient functions and anomalous dimensions for $h^{DGLAP}_{q,g}$ are calculated in several orders in $\alpha_s$. They contain 
contributions essential in region $D_A$ but do not contain total resummations at all. 
In contrast, $h_{q,g}$ calculated in DLA involve all-order DL contributions which are leading in region $D_B$ but are inessential in $D_A$.  
It seems scarcely possible to modify the IREE technology so as to 
include non-leading contributions. It leaves us with the only possibility to describe the parton helicities in region $D_A \oplus D_B$: to obtain an interpolation 
formulae which should coincide with the 
DGLAP ones at large $x, Q^2$ and at the same time it should account for the total resummation of DL contributions. 

\subsection{Helicities in the region $D_A \oplus D_B$  }

Expression for $h_{q,g}$ in the region $D_B$ are given in Eq.~(\ref{hmelcol}). 
In order to adapt them to the region $D_A \oplus D_B$, 
they should be modified so as to combine both the resummation of DL terms and
DGLAP non-DL terms 
in the coefficient functions and anomalous dimensions. In order to make such formulae, the following steps should be done:
\\
\textbf{(i):} subtract DL terms from $C_{i}^{DGLAP}$ and $\Omega_{(\pm)}^{DGLAP}$. We denote the result
$\bar{C}_{i}^{DGLAP}$ and $\bar{\Omega}_{(\pm)}^{DGLAP}$. \\
\textbf{(ii):} Add $\bar{C}_{i}^{DGLAP}$ and $\bar{\Omega}_{(\pm)}^{DGLAP}$ to the expressions for $C_{i}$
and $\Omega_{(\pm)}$ defined in Eqs.~(\ref{solc},\ref{solc1},\ref{omegapm}) respectively:

\begin{eqnarray}\label{regab}
\widetilde{\Omega}_{(\pm)} &=& \Omega_{(\pm)} + \widetilde{\Omega}^{DGLAP}_{(\pm)},
\\ \nonumber
\widetilde{C}_{i} &=& C_{i} + \widetilde{C}^{DGLAP}_{i}.
\end{eqnarray}

Replacing $C_{i}$ and $\Omega_{(\pm)}$ in Eqs.~(\ref{solc},\ref{solc1},\ref{omegapm}) by $\widetilde{C}_{i}$ and $\widetilde{\Omega}_{(\pm)}$,
we obtain interpolating expressions, which we denote $\psi^{AB}_{ij}$ for the helicities, which is valid
in the large-$Q^2$ region $A \oplus B$.


\subsection{Helicities in the region $D_A\oplus D_B \oplus D_C \oplus D_D$}

It was shown in Ref.~\cite{egtsmallq1,egtsmallq2} that extension of the large-$Q^2$ expressions $\psi^{AB}_{ij}$ to
the small-$Q^2$ region $D_C \oplus D_D$ can be done with the replacement of $Q^2$ by $\bar{Q}^2$:

\begin{equation}\label{barq}
\widetilde{Q}^2 = Q^2 + \mu^2
\end{equation}
and therefore $x$ is replaced by $\bar{x}$:

\begin{equation}\label{barx}
\widetilde{x} = x + \mu^2/2pq,
\end{equation}
where $\mu$ is the IR cut-off.
It is worth reminding that shift of $Q^2$ was
suggested in Refs.~\cite{nacht,bad1,bad2,bad3} where the parameter $\mu$ of the shift was introduced on phenomenological basis whereas
it was proved in  Ref.~\cite{egtsmallq1,egtsmallq2} that $\mu$ should be the IR cut-off.
As a result, replacements of $C_j, \Omega_{(\pm)}$
by $\widetilde{C}, \widetilde{\Omega}_{(\pm)}$ followed by replacement of $x, Q^2$ by $\widetilde{x},\widetilde{Q}^2$
convert $\psi_{ik}$ into $\widetilde{\psi}_{ik}$ which can be used in
the whole region $D_A\oplus D_B \oplus D_C \oplus D_D$. Convoluting $\widetilde{\psi}_{ik}$ with the initial parton distributions, we arrive at the expressions
for the parton helicities $\widetilde{\psi}_{q,g}$ valid at arbitrary $Q^2$ and arbitrary $x$:

\begin{eqnarray}\label{factmellinbar}
 \widetilde{\psi}_q \left(\omega, y\right) &=& \widetilde{\psi}_{qq} \left(\omega, y\right) \varphi_q (\omega) + \widetilde{\psi}_{qg} \left(\omega, y\right) \varphi_g (\omega),
 \\ \nonumber
 \widetilde{\psi}_g \left(\omega, y\right) &=& \widetilde{\psi}_{gq} \left(\omega, y\right) \varphi_q (\omega) + \widetilde{\psi}_{gg} \left(\omega, y\right) \varphi_g (\omega).
\end{eqnarray}

In the next Sect. we will show that the inputs $\phi_{q,g}$ can be approximated by constants:

\begin{equation}\label{phin}
\phi_q \approx N_q,~~~\phi_g \approx N_g
\end{equation}
and therefore $h_{q,g}$ in the region $D_{tot}$ can be represented as follows:

\begin{eqnarray}\label{htot}
h_q &=& \int_{-\imath \infty}^{\imath \infty} \frac{d \omega}{2 \pi \imath} \widetilde{x}^{- \omega}
\left[ \widetilde{\psi}_{qq} \left(\omega, \widetilde{y}\right) N_q + \widetilde{\psi}_{qg}  \left(\omega, \widetilde{y}\right) N_g\right],
 \\ \nonumber
h_g &=& \int_{-\imath \infty}^{\imath \infty} \frac{d \omega}{2 \pi \imath} \widetilde{x}^{- \omega}
\left[ \widetilde{\psi}_{gq} \left(\omega, \widetilde{y}\right) N_q + \widetilde{\psi}_{gg}  \left(\omega, \widetilde{y}\right) N_g\right],
\end{eqnarray}
where $\widetilde{x}$ is defined in Eq.~(\ref{barx}) while $\widetilde{\psi}_{ij} \left(\omega, \widetilde{y}\right)$ 
correspond to the expressions in  Eq.~(\ref{solpsi1h},\ref{solpsi2h}) with replacements of
Eqs.~(\ref{regab},\ref{barq}). The non-perturbative factors $N_{q,g}$ cannot be calculated in the QCD framework and should be specified by using other ways, for example by   
fitting experimental data.  

\section{Small-$x$ asymptotics of the parton helicities}

Small-$x$ asymptotics of $h_q$ and $h_g$  are the same save pre-exponential 
factors both in Collinear and $KT$ forms of Factorizations, so we consider below $h_q$ 
in Collinear Factorization. First we consider the small-$x$ asymptotics of $h_q$  in DLA, then in 
the DGLAP approach and compare them. Quite often in the literature, the high-energy asymptotics are simply represented by the factor $s^{\omega_0}$ 
(where $\omega_0$ is the rightmost/leading singularity) without any derivation. However, we find it useful to derive the asymptotics 
from the parent expressions for the helicities, applying  
the Saddle-Point Method, see also the recent paper Ref.~\cite{kovchsaddle}. 

\subsection{Small-$x$ asymptotics of $h_{q,g}$  in DLA}

Expressions of Eq.~(\ref{omegapm}) for $\Omega_{(\pm)}$  read that 
$\Omega_{(+)} > \Omega_{(-)}$, so we neglect the terms $\sim \Omega_{(-)}$ in (\ref{omegapm}). Then 

\begin{eqnarray}\label{homegaplus}
h_q \left(x, y_1\right) \approx \int_{-\imath \infty}^{\imath \infty} \frac{d \omega}{2 \pi \imath} x^{-\omega}
\;\omega\;
 \left[C_{qq}^{(+)}(\omega) \varphi_q (\omega) + C_{qg}^{(+)}(\omega)\varphi_g (\omega)\right]
 e^{y_1 \Omega_{+}(\omega)}, 
  \\ \nonumber 
h_g \left(x, y_1\right) \approx \int_{-\imath \infty}^{\imath \infty} \frac{d \omega}{2 \pi \imath} x^{-\omega}
\;\omega\;\left[C_{gq}^{(+)}(\omega) \varphi_q (\omega) + C_{gg}^{(+)}(\omega)\varphi_g (\omega)\right]
 e^{y_1 \Omega_{+}(\omega)}, 
 \end{eqnarray}
where we denote $C_{ij}^{(+)}$ the factors at the exponential $e^{y \Omega_{+}(\omega)}$ in Eqs.~(\ref{solpsi1h},\ref{solpsi2h}). 
 It is convenient to write Eq.~(\ref{homegaplus}) in such a way: 

\begin{eqnarray}\label{hqge}
h_q \left(x, y_1\right) \approx \int_{-\imath \infty}^{\imath \infty} \frac{d \omega}{2 \pi \imath} 
e^{\omega \xi + E_q(\omega) + y_1\Omega_{(+)}(\omega)},
  \\ \nonumber 
 h_g \left(x, y_1\right) \approx \int_{-\imath \infty}^{\imath \infty} \frac{d \omega}{2 \pi \imath} 
e^{\omega \xi + E_g(\omega) + y_1 \Omega_{(+)}(\omega)},
\end{eqnarray} 
with $\xi = \ln (1/x)$ and 

\begin{eqnarray}\label{eqg}
E_q(\omega) = \ln \left[C_{qq}^{(+)}(\omega) \varphi_q (\omega) + C_{qg}^{(+)}(\omega)\varphi_g (\omega)\right],
\\ \nonumber
E_g(\omega) = \ln \left[C_{gq}^{(+)}(\omega) \varphi_q (\omega) + C_{gg}^{(+)}(\omega)\varphi_g (\omega)\right]
\end{eqnarray}

and apply the Saddle Point method to Eq.~(\ref{homegaplus}). Dealing with the both integrals in Eq.~(\ref{homegaplus}) 
is identical, so we focus on $h_q$. The first step is to expand the exponent in the first equation in Eq.~(\ref{homegaplus}) in the power series, 
retaining three terms:

\begin{eqnarray}\label{hser}
\omega  \rho + E_q(\omega) +y_1 \Omega_{(+)}(\omega)  \approx \left[\omega_0  \rho + E_q(\omega_0) + y_1 \Omega_{(+)}(\omega_0)\right]  
\\ \nonumber
+ 
\left[\xi + E^{\prime}_q(\omega_0) + y_1 \Omega^{\prime}_{(+)}(\omega)\right] (\omega - \omega_0) + 
\frac{1}{2} \left[E^{\prime \prime}_q(\omega_0) + y_1 \Omega^{\prime \prime}_{(+)}(\omega)\right] (\omega - \omega_0)^2,
\end{eqnarray}
with the prime signs denoting derivatives $d/d \omega$ and  $\omega_0$ being the stationary point which is found from the stationary equation: 

\begin{equation}\label{eqomega0}
\xi + E^{\prime}_q(\omega_0) + y_1 \Omega^{\prime}_{(+)}(\omega) =0
\end{equation}
at $\xi \to \infty$. Combining Eqs.~(\ref{hqge},\ref{hser},\ref{eqomega0}), obtain 

\begin{eqnarray}\label{ashgen}
h_{q} \left(x, y_1\right) &\approx&  
e^{-\omega_0 \xi + E_q(\omega_0) + y_1\Omega_{(+)}(\omega_0)}
\int_{-\imath \infty}^{\imath \infty} \frac{d \omega}{2 \pi \imath}
\; e^{\left[(1/2) \left(E^{\prime \prime}_{q}(\omega_0) + y_1 
\Omega_{(+)}^{\prime \prime}(\omega_0)\right) (\omega - \omega_0)^2 \right]}
\\ \nonumber 
&= & \Pi_q (\omega_0)\; \xi^{-3/2}\;e^{-\omega_0 \xi  + y_1\Omega_{(+)}(\omega_0)}, 
\\ \nonumber
h_g\left(x, y_1\right) &\approx&  \Pi_g (\omega_0) \; \xi^{-3/2}\;e^{-\omega_0 \xi  + y_1\Omega_{(+)}(\omega_0)},
\end{eqnarray}
with 

\begin{eqnarray}\label{pi}
\Pi_q (\omega_0) = \left[C_{qq}^{(+)}(\omega_0) \varphi_q (\omega_0) + C_{qg}^{(+)}(\omega_0)\varphi_g (\omega_0)\right]
\sqrt{\frac{2 \pi}{E^{\prime \prime}_q (\omega_0) + y_1 \Omega_{(+)}^{\prime \prime} (\omega_0)}},
\\ \nonumber 
\Pi_g (\omega_0) = \left[C_{gq}^{(+)}(\omega_0) \varphi_q (\omega_0) + C_{gg}^{(+)}(\omega_0)\varphi_g (\omega_0)\right]
\sqrt{\frac{2 \pi}{E^{\prime \prime}_q (\omega_0) + y_1 \Omega_{(+)}^{\prime \prime} (\omega_0)}} .
\end{eqnarray}

Now consider Eq.~(\ref{eqomega0}) in more detail. Obviously, it holds at any $Q^2$ when 

\begin{equation}\label{esing}
E^{\prime}_{q,g}(\omega) \to - \infty
\end{equation}
at $\omega \to \omega_0$. 
So, $E^{\prime}_g(\omega_0)$ 
should be singular 
at $\omega \to \omega_0$. 
Basically, there can be many singularities in expressions in  Eq.~(\ref{eqomega0})  
but we need to find out the leading one. We assume that the fits $\varphi_{q,g} (\omega)$ are 
regular in $\omega$ and, remembering that the factors $C^{(+)}_{q,g}$ 
are made out of $h_{ij}$, 
represent the derivatives $E^{\prime}_{q}(\omega)$ 
as follows: 

\begin{eqnarray}\label{eprime}
E^{\prime}_{q}(\omega) &\approx& \frac{\partial E_q}{\partial h_{ij}} \frac{\partial h_{ij}}{\partial Z}
\left[ \frac{\partial Z}{\partial U} \frac{1}{2 \sqrt{U - \sqrt{V}}}\frac{d U}{d \omega} - \frac{\partial Z}{\partial V} \frac{1}{2 \sqrt{V}}
\frac{d V}{d \omega}\right],
\end{eqnarray}
where we have retained only the terms which can be singular.
 Eq.~(\ref{eprime}) demonstrates that the leading singularity $\omega_0$ can come either from  
 $\sqrt{U - \sqrt{V}} = 0$  or from 
 $\sqrt{V} = 0$, both of them are 
 square-root branching points. However, $1/\sqrt{U - \sqrt{V}} \to + \infty$ and therefore it cannot equate $\rho$, so 
 the only option for the leading singularity corresponds to $V = 0$, 
 in the term $- 1/\sqrt{V}$ which $\to - \infty$ at $\omega \to \omega_0$. It means that the last term in brackets in 
 Eq.~(\ref{eprime}) is the most important and therefore

 \begin{equation}\label{e2prime}
 E^{\prime \prime}_{q}(\omega) \approx - \frac{\partial E_q}{\partial h_{ij}} \frac{\partial h_{ij}}{\partial Z}
\frac{d V}{d \omega} \frac{d}{d \omega} \left(\frac{1}{2 \sqrt{V}}\right) = 
\frac{\partial E_q}{\partial h_{ij}} \frac{\partial h_{ij}}{\partial Z}
\frac{1}{4 V^{3/2}} \left(\frac{d V}{d \omega}\right)^2 .
 \end{equation}
 
  It explains 
 appearance of the factors $\xi^{-3/2}$ in Eq.~(\ref{ashgen}). Then, we remind that the structure function $g_1$ is also made out of 
 amplitudes $h_{ik}$, so the equation for the leading singularity for $g_1$ also includes the expression in 
 brackets of Eq.~(\ref{eprime}) and leads to the requirement $V = 0$. It proves that the asymptotics of 
 the helicities and $g_1$ are identical save the factors $\Pi_i$: 
 
 \begin{equation}\label{as}
 h_{q} \sim h_g \sim g_1 \sim \left[\ln^{-3/2}\left(Q^2/\mu^2\right)\right]\; x^{-\omega_0} \left(Q^2/\mu^2\right)^{\omega_0/2},
 \end{equation}
with $\omega_0$ being the intercept. Obviously, $\omega_0$ does not depend on $Q^2$. When the running coupling effects are accounted for, the intercept $\omega_0 = 0.86$ at $\mu = 1$~GeV, see  Ref.~\cite{egtg1sum} for more detail. The identity 
 of the asymptotics of $h_{q,g}$ and $g_1$ in Eq.~(\ref{as}) without using the Saddle-Point method was 
 obtained in Refs.~\cite{kovchfirst} - \cite{smalldis}.  
 
The square-root singularity $1/\sqrt{V} = 0$ is the result of total resummation of DL contributions 
The perturbation series of DL contributions  to a QCD amplitude $M$
can generically be written in the $\omega$-space as follows (we skip here the $y$-dependence for simplicity): 

\begin{equation}\label{dlser}
M = c_0 + c_1\; \frac{\alpha_s}{\omega^2} + c_2\; \frac{\alpha^2_s}{\omega^4} + c_3\; \frac{\alpha^3_s}{\omega^6} +...,
\end{equation}
where $c_i$, with $i = 1,2,..$, are numerical factors. Although each term in Eq.~(\ref{dlser}) contains a pole singularity at $\omega = 0$, 
the total sum of them forms a more complicated singularity (e.g. the square-root one) and as a result leads to 
the Regge asymptotics. In contrast, 
the truncated sum of DL terms in (\ref{dlser}) never leads to asymptotics of the Regge type. We prove it, discussing the DGLAP 
small-$x$ asymptotics. Besides, one more difference between the asymptotics obtained in our approach and DGLAP is that the intercept in 
our approach is equally generated by the coefficient functions and the anomalous dimensions while the coefficient 
function impact is negligibly small in DGLAP.   

\subsection{DGLAP small-$x$ asymptotics}

In this Sect. we discuss the small-$x$ asymptotics generated by the perturbative components of DGLAP\cite{dglap1,dglap2,dglap3,dglap4}, 
postponing consideration of an influence of the fits for parton distributions until the next Sect.  
Consider some structure function/helicity $F$ in the framework DGLAP, assuming that the fits for initial parton 
distributions do not have singularities in $x$, so their impact can be neglected. Represent $F$ generically as follows 
(cf. Eq.~(\ref{mellin})): 

\begin{eqnarray}\label{fdglap}
F &=& \int_{- \imath \infty}^{ \imath \infty} \frac{d \omega}{2 \pi \imath} e^{\omega \xi + \ln C(\omega) + y h(\omega)},
\end{eqnarray}
where $C(\omega)$ ($h(\omega)$) corresponds to the coefficient functions (anomalous dimensions). 
Expand the exponent in the series: 

\begin{eqnarray}\label{fser}
\omega \xi + \ln C(\omega) + y h(\omega) \approx \omega_n \xi + \ln C(\omega_n) + y h(\omega_n) 
\\ \nonumber
+ \left[\xi + (\ln C(\omega_n))^{\prime} + y h^{\prime}(\omega_n) \right] (\omega - \omega_n)
+ \frac{1}{2}\left[(\ln C(\omega_n))^{\prime \prime} + y h^{\prime \prime}(\omega_n) \right] (\omega - \omega_n)^2.
\end{eqnarray}

Equation for $\omega_n$: 

\begin{equation}\label{eqomegangen}
\xi + \ln C(\omega_n))^{\prime} + y h^{\prime}(\omega_n) = 0. 
\end{equation}

DL contributions to $C$ and $h$ in LO, NLO, NNLO, etc. can be represented as follows: 

\begin{eqnarray}\label{chdglap}
C(\omega) &=& 1 + c_1 \frac{\alpha_s}{\omega^2} + c_2 \frac{\alpha^2_s}{\omega^4} + c_3 \frac{\alpha^3_s}{\omega^6} +...+ 
c_n \frac{\alpha^n_s}{\omega^{2n}}
\\ \nonumber
&=& \left[\omega^{2n} + c_1 \alpha_s \omega^{2n-2} + c_2 \alpha^2_s \omega^{2n-4} 
.. + c_n \alpha^n_s\right]\omega^{- 2n},
\\ \nonumber
h(\omega) &=& c^{\prime}_1 \frac{\alpha_s}{\omega} + c^{\prime}_2 \frac{\alpha^2_s}{\omega^3} + 
...+ c^{\prime}_n \frac{\alpha^n_s}{\omega^{2n - 1}}.
\end{eqnarray}

Therefore, 

\begin{equation}\label{lnc}
\ln C = \ln \left[\omega^{2n} + c_1 \alpha_s \omega^{2n-2} + c_2 \alpha^2_s \omega^{2n-4} 
.. + c_n \alpha^n_s\right] - 2n \ln \omega
\end{equation}
and 
\begin{eqnarray}\label{lncprime}
\left(\ln C(\omega)\right)^{\prime} &=&   
\frac{\left[2n \omega^{2n} + c_1 (2n - 2) \alpha_s \omega^{2n-2} + 
.. + c_{n-1} 2 \alpha^{n-1}_s \omega\right]}
{\omega \left[\omega^{2n} + c_1 \alpha_s \omega^{2n-2} 
.. + c_n \alpha^n_s\right]} 
\\ \nonumber
&-& \frac{2n}{\omega}.
\end{eqnarray}

When $\omega \to n$, the singular contribution to $\left(\ln C(\omega)\right)^{\prime}$ is 

\begin{equation}\label{csing}
\left(\ln C(\omega)\right)^{\prime}_{sing}  = - \frac{2n}{\omega},
\end{equation}
while the singular contribution to $h$ is

\begin{equation}\label{hsing}
h^{\prime}_{sing} = - (2n - 1)c^{\prime}_n \frac{\alpha^n_s}{\omega^{2n}}. 
\end{equation}

Substituting Eqs.~(\ref{csing},\ref{hsing}) in Eq.~(\ref{eqomegangen}), obtain

\begin{equation}\label{eqomegan}
\xi  - \frac{2n}{\omega_n}  - y\;(2n - 1)c^{\prime}_n  \frac{\alpha^n_s}{\omega_n^{2n}} \approx 
\xi  - y\;(2n - 1)c^{\prime}_n \frac{\alpha^n_s}{\omega_n^{2n}} = 0,
\end{equation}
therefore 

\begin{equation}\label{omegan}
\omega_n = \left(y\; (2n- 1)c^{\prime}_n \alpha^n_s/\xi\right)^{1/2n},
\end{equation}
which leads to the following asymptotics of $F$:

\begin{equation}\label{asdglap}
F \sim e^{\xi \omega_n} = \exp \left[\xi^{(1 - 1/2n)} y^{1/2n} \left((2n-1)\alpha^n_s c^{\prime}_n \right)^{1/2n}\right].
\end{equation}

In particular, putting $n=1$ leads to the well-known formula for asymptotics of LO DGLAP:  $F_{LO} \sim \exp \left[c_{LO}\sqrt{\alpha_s \xi y}\right]$, 
whereas the asymptotics of NLO DGLAP corresponds to $n = 2$: $F_{NNLO} \sim \exp \left[c_{NLO}\xi^{3/4} y^{1/4} 
\sqrt{\alpha}\right]$.\\

\textbf{Remark on the false DGLAP intercept}  \\

Sometimes\footnote{For instance, when DGLAP is combined with BFKL}, when DGLAP applies to experimental data processing, the DGLAP asymptotics (\ref{asdglap}) 
are interpreted in  the Regge-like way: Neglecting the term $1/2n$ in the exponential $\xi^{1 - 1/2n}$, they approximate  Eq.~(\ref{asdglap}) by the Regge-like 
expression 

\begin{equation}\label{dglapregge}
F \sim \xi^{\Delta}, 
\end{equation}
where the false "intercept" 

\begin{equation}\label{deltadglap}
\Delta =  \sqrt{\alpha_s}\;C\;y^{1/2n}
\end{equation}
depends on $Q^2$ and $n$ corresponds to N$^n$LO DGLAP while $C$ is a numerical factor.  Obviously, $\Delta$ contradicts both 
the phenomenological 
Regge theory and the 
asymptotics (\ref{as}). The crucial difference between them is $Q^2$-dependence of $\Delta$ whereas the $Q^2$-dependence of the small-$x$ asymptotics in 
Eq.~(\ref{as})  
comes from the factor $(Q^2/\mu^2)^{\omega_0/2}$. Obviously, the $Q^2$-dependence predicted by Eq.~(\ref{as}) is much sharper than the one in 
Eq.~(\ref{deltadglap}). Which is those predictions is true,  can be checked with analysis of  
experimental data.

\subsection{Asymptotics induced by DGLAP fits for initial parton densities}

Adding to Eq.~(\ref{fdglap}) the fit for initial parton densities converts Eq.~(\ref{fdglap}) into 

\begin{eqnarray}\label{fdglapfit}
F &=& \int_{- \imath \infty}^{ \imath \infty} \frac{d \omega}{2 \pi \imath} e^{\omega \xi + \ln C(\omega) + y h(\omega)} \varphi (\omega),
\end{eqnarray}
where $\varphi (\omega)$ stands for the fit. 
Typical DGLAP fits for the parton densities have the following generic structure:

\begin{eqnarray}\label{fit}
\Phi &=&  N x^{-a} (1 - x)^b (1 + c x^d),
\end{eqnarray}
where parameters $N,a,b,c,d$ are positive. Expand $\Phi$ into the power series:

\begin{eqnarray}\label{fitser}
\Phi &=&   N \left[ \sum_{k = 0}^{\infty} \lambda_k^b\; x^{-a +k}
+ c \sum_{k = 0}^{\infty} \lambda_k^b\; x^{-a + d +k}\right],
\end{eqnarray}
where $\lambda_k^b$ are binomial factors. From the point of the Regge theory, the expansion in Eq.~(\ref{fitser}) is
the series of Reggeons.
In the $\omega$-space it converts into the series of the simple poles:

\begin{eqnarray}\label{fitomega}
\varphi (\omega) &=& N \left[\sum_{k = 0}^{\infty} \frac{\lambda^b_k}{\omega - a + k}
+ c \sum_{k = 0}^{\infty} \frac{\lambda^b_k}{\omega - a + d + k}\right],
\end{eqnarray}
with the pole 

\begin{equation}\label{leadpole}
\varphi_0 (\omega) = \frac{N}{\omega - a}
\end{equation}
being leading.  Substituting $\varphi_0$ in Eq.~(\ref{fdglapfit}), obtain the small-$x$ asymptotics of $F$ 
of the Regge kind:

\begin{eqnarray}\label{fa}
F \sim  \left[N C(a)\right]\;  x^{-a}\; \left(Q^2/\mu^2\right)^{h (a)},
\end{eqnarray}
whereas the other poles in Eq.~(\ref{fitomega}) bring sub-leading corrections. It is interesting to notice that the asymptotics (\ref{fa}) and (\ref{as}), in contrast to Eq.~(\ref{deltadglap}),  
predict the Regge kind of $x$- dependence with intercepts independent of $Q^2$ .

\section{Conclusions and outlook}

In this work, we have obtained explicit expressions for the  parton helicities $h_{q,g} (x, Q^2)$ at high energies, which can be used in the framework of both 
Collinear and $KT$ forms of QCD Factorization. 
It was done in three steps: \\
Firstly, we calculated the perturbative components $M_{ij}$ of the helicities in the DLA 
by constructing and solving appropriate IREEs. This  
allowed us to obtain $h_{q,g}(x, Q^2)$ in the kinematic region of small $x$ and large $Q^2$. 
We demonstrated both similarity and difference between IREEs for the parton helicities and for the DIS structure function $g_1$. Also, we 
considered the color octets impact on the helicities. \\
Secondly,  we obtained interpolation expressions for $h_{q,g}(x, Q^2)$ which hold at arbitrary $x$ and large  $Q^2$. It was done by 
combining the appropriate 
DGLAP formulae with the formulae obtained in DLA. By doing so, we obtained the universal expressions for $h_{q,g}(x, Q^2)$, which on one hand coincide with the DGLAP 
expressions at large and medium $x$ and on the other hand contain total resummations of  DL contributions important at small $x$. \\
Thirdly, we extended our formalism so as to include in consideration the region of small $Q^2$. It was done with the shift of $Q^2$ introduced in Eq.~(\ref{barq}). 
As a result, we arrived at the expressions for $h_{q,g}(x, Q^2)$ valid at arbitrary $x$ and $Q^2$. They can be used in both Collinear and KT Factorization 
context. 

After obtaining the expressions for $h_{q,g}(x, Q^2)$, we considered in detail the topic usually discussed too briefly in the literature, namely the 
small-$x$ asymptotics of the helicities. As the asymptotics  in Collinear and $KT$  forms of QCD Factorization 
are quite similar, we focused on Collinear Factorization. 
We demonstrated that though $h_q(x, Q^2)$, $h_g(x, Q^2)$, and the structure function $g_1(x, Q^2)$ are represented in DLA by different formulae, their small-$x$ asymptotics 
are identical.   Then we compared the small-$x$ asymptotics of $h_{q,g}(x, Q^2)$ in DLA and 
DGLAP. We showed that neither LO DGLAP nor N$^n$LO DGLAP can provide the helicities with the Regge asymptotics whereas 
 the factors $x^{-a}$ in the DGLAP fits for initial 
parton densities can make it.  

Discussing $KT$ Factorization, we argued that this form of QCD Factorization is the only self-consisting means 
for including in consideration the parton Orbital Angular Momenta (OAM) when the longitudinal nucleon spin is discussed 
whereas applying Collinear Factorization is inconsistent from the theory considerations. 
As a consequence, results of the papers combining   OAM and Collinear Factorization (e.g. through using DGLAP) 
should be revised. 
It is worth mentioning that Refs.~\cite{oam1,oam2,oam3}, where the OAM  contribution to the proton spin was suggested, as well as the
overview\cite{leader} do not relate OAM to Collinear Factorization.
On the other hand, we found out that involving OAM to description of the transverse spin of nucleons is compatible with Collinear Factorization. To conclude this subject we stress  
once more that we agree   
that  accounting for OAM is the interesting and important subject but we argue against combining Collinear Factorization and OAM.


\begin{thebibliography}{99}






\bibitem{kovchfirst}  Yuri V. Kovchegov, Daniel Pitonyak, Matthew D. Sievert.
JHEP 01 (2016) 072; JHEP 10 (2016) 148 (erratum).

\bibitem{kovchns} Yuri V. Kovchegov, Daniel Pitonyak, Matthew D. Sivert.
Phys. Rev. D 95 (2017) 1, 014033.

\bibitem{disagr1} Yuri V. Kovchegov, Daniel Pitonyak, Matthew D. Sivert.  Phys. Rev. Lett. 118 (2017) 5, 052001.

\bibitem{kovch3} Y. V. Kovchegov, D. Pitonyak and M. D. Sievert, JHEP 10 (2017) 198.


\bibitem{kovch2} Y. V. Kovchegov and M. D. Sievert. Phys. Rev. D 99 (2019) 054032.

\bibitem{improv}    Florian Cougoulic, Yuri V. Kovchegov. Phys.Rev.D 100 (2019) 11, 114020.

\bibitem{improv2}    Yuri V. Kovchegov, Andrey Tarasov, Yossathorn Tawabutr. JHEP 03 (2022) 184.

\bibitem{improv3} F. Cougoulic, Y. V. Kovchegov, A. Tarasov and Y. Tawabutr. JHEP 07 (2022) 095.

\bibitem{smalldis}    Jeremy Borden, Yuri V. Kovchegov. Phys. Rev. D 108 (2023) 1, 014001. 


\bibitem{berns} J.~Bartels, B.I.~Ermolaev, M.G.~Ryskin. Z.Phys.C 72 (1996) 627.
\bibitem{bers} J.~Bartels, B.I.~Ermolaev, M.G.~Ryskin. Z.Phys.C 70 (1996) 273.

\bibitem{agr}Florian Cougoulic, Yuri V. Kovchegov, Andrey Tarasov and Yossathorn Tawabutr.
JHEP 07 (2022) 095.


\bibitem{adam}   Daniel Adamiak, Yuri V. Kovchegov and Yossathorn Tawabutr.
Phys.Rev.D 108 (2023) 5, 5.

\bibitem{oam} Yuri V. Kovchegov and Brandon Manley. JHEP 02 (2024) 060; JHEP 08 (2024) 140 (erratum). 

\bibitem{hatta}     Renaud Boussarie, Yoshitaka Hatta, Feng Yuan.
Phys. Lett. B 797 (2019) 134817.

\bibitem{taw} Yossathorn Tawabutr.
eprint 2311.18185 [hep-ph].

\bibitem{global} Daniel Adamiak, Nicholas Baldonado, Yuri V. Kovchegov, W. Melnitchouk, Daniel Pitonyak,
Nobuo Sato, Matthew D. Sievert, Andrey Tarasov, and Yossathorn Tawabutr.
Phys. Rev. D 108 (2023) 11, 11.

\bibitem{kovscatt} D.~Adamiak, N.~Baldonado, Yuri V.~Kovchegov, Ming Li, W.~Melnitchouk. Arxiv: 2503.21006.

\bibitem{espin} B.I. Ermolaev.     Eur.Phys.J.C 85 (2025) 3, 360. 

\bibitem{rhic1} E. C. Aschenauer et al. Arxiv: 1304.0079.

\bibitem{rhic2} E.-C. Aschenauer et al. Arxiv: 1501.01220.

\bibitem{egtg1s} B.I.~Ermolaev, M.~Greco, S.I.~Troyan. Phys. Lett. B 579 (2004) 321.

\bibitem{egtg1sum} B.I.~Ermolaev, M.~Greco, S.I.~Troyan. Nuovo Cim. 33 (2010) 2, 57.



\bibitem{g} V.N.~Gribov. Sov.J.Nucl.Phys. 5 (1967) 280. 

\bibitem{l1} L.N.~Lipatov. Zh.Eksp.Teor.Fiz.82 (1982)991.

\bibitem{l2} L.N.~Lipatov. Phys.Lett.B116 (1982)411.

\bibitem{kl1} R. Kirschner and L.N. Lipatov. ZhETP 83(1982)488.

\bibitem{kl2} R. Kirschner and L.N. Lipatov.  Nucl. Phys. B 213(1983)122.

\bibitem{efl} B.I.~Ermolaev, V.S.~Fadin, L.N.~Lipatov. Yad.Fiz. 45 (1987) 817-823. 

\bibitem{el} B.I.~Ermolaev, L.N.~Lipatov. Int.J.Mod.Phys.A 4 (1989) 3147.

\bibitem{ggfl} V.G.~Gorshkov, V.N.~Gribov, G.V.~Frolov, L.N.~Lipatov. Phys.Lett. 22 (1966) 671.

\bibitem{dglap1}
G.~Altarelli and G.~Parisi, Nucl.~Phys.B126 (1977) 297. 

\bibitem{dglap2} V.N.~Gribov and L.N.~Lipatov, Sov.~J.~Nucl.~Phys. 15 (1972) 438.

\bibitem{dglap3} L.N.Lipatov, Sov.~J.~Nucl.~Phys. 20 (1972) 95.  

\bibitem{dglap4} Yu.L.~Dokshitzer, Sov.~Phys.~JETP 46 (1977) 641.


\bibitem{pms} P.M.~Stevenson. Phys.Rev.D 23 (1981) 2916.


\bibitem{oam1} John R. Ellis, Marek Karliner.  Phys.Lett.B 213 (1988) 73. 

\bibitem{oam2} R. L. Jaffe and A. Manohar, Nucl. Phys. B 337 (1990) 509.

\bibitem{oam3} X.-D. Ji. Phys. Rev. Lett. 78 (1997) 610, [hep-ph/9603249].

\bibitem{leader} E.~Leader, C.~Lorce. Phys.Rept. 541 (2014) 3, 163. 

\bibitem{hatta2} S.~Benic, Y.~Hatta. arXiv: 2505.05172.


\bibitem{egtsmallq1} B.I.~Ermolaev, M.~Greco, S.I.~Troyan. Eur. Phys. J. C50(2007)823; 

\bibitem{egtsmallq2} B.I.~Ermolaev, M.~Greco, S.I.~Troyan. Eur. Phys. J.
C51(2007)859.

\bibitem{nacht} O.Nachtmann. Nucl. Phys. B 63 (1973) 237. 

\bibitem{bad1} B.~Badelek and J.~Kwiecinski. Z.~Phys. C 43 (1989) 251; 

\bibitem{bad2} B.~Badelek and J.~Kwiecinski. Rev. Mod. Phys. 68 (1996)445. 

\bibitem{bad3} B.~Badelek and J.~Kwiecinski. Phys. Lett. B 418 (1998) 229.

\bibitem{kovchsaddle} Jeremy Borden, Yuri V. Kovchegov. arXiv:250800195.  
























	

\end{thebibliography}
\end{document}